\documentclass[aps,prd,amsmath,floats,floatfix,twocolumn,nofootinbib]{revtex4}
\usepackage{graphicx,amssymb,amsmath,amsbsy,mathrsfs}
\usepackage{bm}
\usepackage[usenames]{color}

\begin{document}
\vspace{-2.5cm} 

\title{Simulations of Binary Black Hole Mergers Using Spectral Methods}

\author{B\'ela Szil\'agyi, Lee Lindblom, and Mark A. Scheel}
\affiliation{Theoretical Astrophysics 350-17,
    California Institute of Technology, Pasadena, CA 91125}

\date{\today}

\begin{abstract}
Several improvements in numerical methods and gauge choice are
presented that make it possible now to perform simulations of the
merger and ringdown phases of ``generic'' binary black-hole evolutions
using the pseudo-spectral evolution code SpEC.  These improvements
include the use of a new damped-wave gauge condition, a new grid
structure with appropriate filtering that improves stability, and
better adaptivity in conforming the grid structures to the shapes and
sizes of the black holes.  Simulations illustrating the success of
these new methods are presented for a variety of binary black-hole
systems.  These include fairly ``generic'' systems with unequal masses
(up to 2:1 mass ratios), and spins (with magnitudes up to $0.4M^2$)
pointing in various directions.
\end{abstract}

\maketitle


\section{Introduction}
\label{s:Introduction}

Black-hole science took a great stride forward in 2005 when
Pretorius~\cite{Pretorius2005a} performed the first successful full
non-linear dynamical numerical simulation of the inspiral, merger, and
ringdown of an orbiting black-hole binary system; this initial success
then stimulated other groups to match this achievement within
months~\cite{Campanelli2006a, Baker2006a}.  These developments lead
quickly to advances in our understanding of black-hole physics:
investigations of the orbital mechanics of spinning
binaries~\cite{Campanelli2006c, Campanelli2006d, Campanelli2007b,
  Herrmann2007c, MarronettiEtAl:2008, Berti2007b}, studies of the
recoil from the merger of unequal mass binary
systems~\cite{Campanelli2005, Herrmann2007b, Baker2006c,
  Gonzalez2007}, the remarkable discovery of unexpectedly large recoil
velocities from the merger of certain spinning binary
systems~\cite{Herrmann2007c, Campanelli2007, Gonzalez2007b,
  Bruegmann-Gonzalez-Hannam-etal:2007, Herrmann2007,
  Choi-Kelly-Boggs-etal:2007, Baker2007, Tichy:2007hk, Schnittman2007,
  Campanelli2007a, Koppitz2007, MillerMatzner2008, Baker2008,
  Healy2008}, investigations into the mapping between the binary
black-hole initial conditions (individual masses and spins) and the
final state of the merged black hole~\cite{Boyle2007a,Boyle2007b,
  Buonanno2008a,Tichy2008,Kesden2008,Rezzolla2008,Barausse2009,Lousto2009},
and improvements in our understanding of the validity of approximate
binary black-hole orbital calculations using post-Newtonian
methods~\cite{Buonanno-Cook-Pretorius:2007, Baker2006d,Pan2007,
  Buonanno2007, Hannam2007,Gopakumar:2007vh, Hannam2007c,
  Hinder2008b}.

These first results on binary black-hole systems were obtained by
several different groups using different codes based on two different
formulations of the Einstein equations (BSSN and generalized
harmonic), using two different methods for treating the black-hole
interiors (moving puncture and excision), and using rather different
gauge conditions to fix the spacetime coordinates.  All of these early
results, however, were obtained with codes based on finite difference
numerical methods and adaptive mesh refinement for computational
efficiency.  The Caltech/Cornell collaboration decided a number of
years ago to follow a different path by developing an Einstein
evolution code, called SpEC,
based on spectral methods.  The advantages of spectral
methods are their superior accuracy and computational efficiency, and
their extremely low numerical dissipation that is ideal for solving
wave propagation problems with high precision.  The disadvantages are
the relative complexity of spectral codes, the lack of
any appropriate pre-existing
spectral code infrastructure (analagous to 
CACTUS~\cite{CACTUS} for example), 
and the extreme sensitivity of spectral algorithms
to developing instabilities when any aspect
of the solution method is mathematicaly ill-posed.  Consequently it 
has taken  our group
somewhat longer to bring  SpEC up to the level of the 
state-of-the-art codes in the field.  

We, along with our Caltech/Cornell collaborators,
have developed a large number of numerical and analytical 
tools over the years that make it possible for  SpEC
to evolve binary black-hole systems with
greater precision 
than any other code (at the present time)~\cite{Hannam:2009hh}.
These technical developments include the
derivation and implementation of constraint preserving and physical
(no incoming gravitational wave) boundary
conditions~\cite{Lindblom2006,Rinne2007}, dual frame evolution methods
and feedback and control systems to lock the computational grids onto
the location of the black holes~\cite{Scheel2006}, and special angular
filtering methods needed to cure an instability  that occurs
when tensor fields are evolved~\cite{Kidder2005}.  Using these methods
we and our collaborators
have performed a number of high precision evolutions of the
inspiral portions of binary black-hole systems, and have used the
gravitational waveforms from these evolutions to calibrate the
accuracy of the approximate post-Newtonian waveforms that are widely
used in gravitational wave data analysis~\cite{Boyle2007,Boyle2008a,
  Boyle2008b, Buonanno:2009qa}.  

During the past year  our group has
begun to have some success in performing
the more dynamical and difficult (for spectral codes) merger and 
ringdown portions of
binary black-hole evolutions~\cite{Scheel2008,Boyle2008a,
Buonanno:2009qa,Lovelace:2009,
  CohenPfeiffer2008, Chu2009}.  The techniques we used to achieve
these first merger calculations turned out to be rather non-robust
however, requiring a great deal of hands-on adjustment and fine tuning
for each new case we attempted.  In this paper we report on several new
numerical and analytical developments that allow us now to perform
stable and accurate binary merger and ringdown simulations of
a fairly wide range of binary black-hole systems.  These new methods
appear to be quite robust, and no fine tuning is required: the same
basic method works on each of the cases we have attempted.

The three new technical breakthroughs that allow us now to perform
successful binary black-hole merger and ringdown simulations are
described in Sec.~\ref{s:TechnicalDevelopments}.  These major advances
include the development of a new gauge condition that works extremely
well with the generalized harmonic formulation of the Einstein system
used in our code.  This new gauge chooses spatial
coordinates that are solutions of a damped wave equation, and chooses
the time coordinate in a way that limits the growth of $\sqrt{g}/N$,
here $g$ is the determinant of the spatial metric and $N$ is the
lapse function.  This new gauge condition is described in some detail
in Sec.~\ref{s:DampedWaveGauge}.  Another important development,
reported in Sec.~\ref{s:GridStructureFiltering}, is the construction of
a new grid structure on which our binary black-hole evolutions are
performed.  This new non-overlapping grid structure, and a certain type of
spectral filtering used with the new grid structure, removes a
class of numerical instabilities that were a limiting factor in our
ability to perform the highly dynamical portions of merger
calculations, and allows us to increase
resolution near the black holes in a more targeted and computationally
efficient way.
The third technical development is a new
more efficient and robust method to conform the structure of the grid
to the shape and size of the black holes in an automatic dynamical way.
These developments are described in Sec.~\ref{s:AdaptiveConformingGrid}
and the Appendix~\ref{s:control-systems}.

Using these new technical tools we have now performed successful
merger and the succeeding ringdown simulations of a fairly wide range
of binary black-hole systems.  We describe in
Sec.~\ref{s:MergerRingdownSimulations} merger simulations for
six reasonably diverse cases.  These include two systems in
which the black holes are non-spinning: one has equal mass black
holes, the other has holes with a 2:1 mass ratio.  We also describe
successful merger and ringdown calculations with two different
equal-mass binary systems in which the holes have identical intrinsic
spins of magnitude $0.4 M^2$: aligned to the orbital angular momentum
in one system, and anti-aligned in the other.  We have also performed
successful merger and ringdown simulations on two fairly generic
binary systems.  These two systems have black holes with the same mass
ratio $M_2/M_1=2$, and  the same ``randomly oriented'' initial
spins of magnitudes $0.2M^2_1$ and $0.4M^2_2$ respectively.  But the
initial separations of the holes are different in the two cases: one
starts about 8.5 orbits before merger, and another (used to perform
careful convergence studies) starts about 1.5 orbits before merger.
The same methods were used to perform all of these merger and ringdown
simulations, and no particular fine tuning was required.

\section{Technical Developments}
\label{s:TechnicalDevelopments}

This section describes the three new technical breakthroughs that
allow us now to perform successful binary black-hole merger and
ringdown simulations.  These major advances include the development of
a new gauge condition, described in Sec.~\ref{s:DampedWaveGauge}, in
which the spatial coordinates satisfy a damped wave equation and the
time coordinate is chosen in a way that controls the
growth of the spatial volume element.  Another
important development, described in
Sec.~\ref{s:GridStructureFiltering}, is a new non-overlapping grid
structure for our binary black-hole simulations, and a type of
spectral filtering for these new grid structures that improves their
accuracy and stability.  The third new technical development,
described in Sec.~\ref{s:AdaptiveConformingGrid}, is a more efficient
and robust method of conforming the structure of the grid to the shape
and size of the black holes in a dynamical and automatic way.

\subsection{Damped-Wave Gauge}
\label{s:DampedWaveGauge}

Harmonic gauge is defined by the condition that each coordinate $x^a$
satisfies the co-variant scalar wave equation:
\begin{eqnarray}
\nabla^c\nabla_c x^a = H^a = 0.
\label{e:HarmonicGauge}
\end{eqnarray}
Harmonic coordinates have proven to be extremely useful for analytical
studies of the Einstein equations, but have found only limited success
in numerical problems like simulations of complicated highly dynamical
black-hole mergers.  One reason for some of these difficulties is
the wealth of ``interesting'' dynamical solutions to the harmonic
gauge condition itself, Eq.~(\ref{e:HarmonicGauge}).  Since all
``physical'' dynamical fields are expressed in terms of the
coordinates, an ideal gauge condition would limit coordinates to those
that are simple, straightforward, dependable, and non-singular; having
``interesting'' dynamics of their own is {\it not} a desirable feature
for coordinates.  The dynamical range available to harmonic
coordinates can be reduced by adding a damping term to the
equation:~\cite{Lindblom2009c,Choptuik09}
\begin{eqnarray}
\nabla^c\nabla_c x^a = \mu_S t^c\partial_c x^a = \mu_S t^a,
\end{eqnarray}
where $t^a$ is the future directed unit normal to the constant-$t$
hypersurfaces.  Adding such a damping term to the equations for the
spatial coordinates $x^i$ tends to remove extraneous gauge dynamics
and drives the coordinates toward solutions of the co-variant spatial
Laplace equation on the timescale $1/\mu$.  Choosing $1/\mu$ to be
comparable to (or smaller than) the characteristic timescale of a
particular problem should remove any extraneous coordinate dynamics on
timescales shorter than the physical timescale. The addition of such a
damping term in the time-coordinate equation is not appropriate
however.  Such a damped-wave time coordinate is driven toward a
constant value, and therefore toward a state in which it fails to be a
useful time coordinate at all. It makes sense then to use the
damped-wave gauge condition only for the spatial coordinates:
\begin{eqnarray}
\nabla^c\nabla_c x^i = H^i = \mu_S t^i = - \mu_S N^i/N,
\label{e:DampedWaveShift}
\end{eqnarray}
where $N^i$ is the shift, and $N$ is the lapse. The appropriate
contra-variant version of this damped-wave gauge condition is
therefore
\begin{eqnarray}
H_a = -\mu_S g_{ai} N^i / N,
\label{e:DampedWaveCoordinates}
\end{eqnarray}
where $g_{ab}$ is the spatial metric of the constant-$t$ hypersurfaces.
(Note that we use
Latin letters from the beginning of the alphabet, $a,b,c,\ldots$,
to denote four-dimensional spacetime indices, and letters
from the middle of the alphabet, $i,j,k,\ldots$,
for three-dimensional spatial indices.)

While the damped-wave gauge is a poor choice for the time coordinate,
the idea of imposing a gauge that adds dissipation to the gauge
dynamics of the time coordinate is attractive.  To find the
appropriate expression for $t^aH_a$, the component of $H_a$ not fixed
by Eq.~(\ref{e:DampedWaveCoordinates}), we note that the gauge
constraint $H_a+\Gamma_a=0$ implies that $t^aH_a$ is given by
\begin{eqnarray}
t^a H_a =t^a\partial_a \log \left(\frac{\sqrt{g}}{N}\right)-
N^{-1}\partial_k N^k,
\label{e:AlternateLapseConstraint}
\end{eqnarray}
where $g=\det g_{ij}$ is the spatial volume element.  In our
experience, a frequent symptom of the failure of simpler gauge
conditions in binary black-hole simulations is an explosive growth of
$g$ in the spacetime region near the black-hole horizons.  This
suggests that a good use of the remaining gauge freedom would be to
attempt to control the growth of the spatial
volume element, $g$.  Choosing the gauge condition,
\begin{eqnarray}
t^a H_a =-\mu_L\log \left(\frac{\sqrt{g}}{N}\right),
\label{e:DampedWaveLapse}
\end{eqnarray}
together with Eq.~(\ref{e:AlternateLapseConstraint}) implies the
following evolution equation for $\sqrt{g}/N$:
\begin{eqnarray}
t^a\partial_a \log \left(\frac{\sqrt{g}}{N}\right)
+\mu_L\log \left(\frac{\sqrt{g}}{N}\right)=
N^{-1}\partial_k N^k,
\label{e:LapseGaugeCondition}
\end{eqnarray}
whose solutions tend to suppress growth in $\sqrt{g}/N$.  The
discussion of this gauge condition in Ref.~\cite{Lindblom2009c} shows
that it also implies that the lapse $N$ satisfys a damped wave
equation, with the damping factor $\mu_L$.  So in this sense, the
gauge condition on the time coordinate, Eq.~(\ref{e:DampedWaveLapse}),
is the natural extension of the spatial-coordinate damped-wave gauge
condition, Eq.~(\ref{e:DampedWaveCoordinates}).

Combining this new lapse condition, Eq.~(\ref{e:DampedWaveLapse}),
with the damped-wave spatial coordinate condition,
Eq.~(\ref{e:DampedWaveCoordinates}), gives the gauge-source
function for our full damped-wave gauge condition:
\begin{eqnarray}
H_a=\mu_L\log\left(\frac{\sqrt{g}}{N}\right)t_a-\mu_S N^{-1}g_{ai}N^i.
\label{e:FullDampedWaveGauge}
\end{eqnarray}
The damping factors $\mu_L\geq0$ and $\mu_S\geq0$ can be chosen quite
arbitrarily as functions of spacetime coordinates $x^a$, or even as
functions of the spacetime metric $\psi_{ab}$.  The gauge source
function $H_a$ depends only on coordinates and the spacetime metric
$\psi_{ab}$ in this case, so these gauge conditions can be implemented
directly in the  generalized harmonic
Einstein system without the need for a gauge
driver.  Previous studies of this condition (developed initially as a
test of the first-order gauge-driver system~\cite{Lindblom2009c})
showed it to be quite useful for evolving single black-hole
spacetimes.  In those tests, which included several different
evolutions of maximal-slice Schwarzschild initial data with large
non-spherical gauge perturbations, the black hole always evolved
quickly toward a non-singular time-independent equilibrium state.

The solutions to the lapse gauge condition,
Eq.~(\ref{e:LapseGaugeCondition}), can be thought of as equating
$\log(\sqrt{g}/N)$ to a certain weighted time average of
$\partial_kN^k/N$.  The timescale associated with this time averaging
is set by $\mu_L$, which determines for example the rate at which
$\sqrt{g}/N$ is driven toward an asymptotic equilibrium
state.  When $\mu_L$ is
constant, it is easy to show that $\sqrt{g}/N$ is driven exponentially
toward this asymptotic state.  
In highly dynamical spacetimes, however, we
find that $\sqrt{g}/N$ must be driven even faster than exponential in
order to prevent the formation of singularities in $g$.  This can be
accomplished by making $\mu_L$ larger whenever $g$ becomes large.  In
practice we find that the choice $\mu_L = \mu_0 [\log(\sqrt{g}/N)]^2$,
(where $\mu_0$ is a constant or perhaps a function of time) is very
effective at suppressing the growth of these singularities.  We find
that choosing the same damping factor $\mu_S$,
\begin{eqnarray}
\label{eq:GaugeDampingFactor}
\mu_S=\mu_L= \mu_0  \left[\log\left(\frac{\sqrt{g}}{N}\right)\right]^2,
\end{eqnarray}
for the spatial part of the gauge condition is also quite effective.
The binary black-hole merger and ringdown simulations described in
Sec.~\ref{s:MergerRingdownSimulations} use these $\mu_L$ and $\mu_S$,
with $\mu_0$ taken to be an order-unity function of time (to
accomodate starting up evolutions from initial data satisfing a
different gauge condition).

\subsection{Grid Structure and Filtering}
\label{s:GridStructureFiltering}

The pseudo-spectral numerical methods used by our code, the Spectral
Einstein Code (SpEC)~\cite{Lindblom2006,Scheel2006,Boyle2006},
represent dynamical fields on a spatial grid structure that is
specially constructed for each problem.  For binary black-hole
simulations,  our group has
been using grid structures constructed from layers
of spherical-shell subdomains centered on each black hole, surrounded
and connected by cylindrical, cylindrical-shell, 
and/or rectangular-block
subdomains, all surrounded by spherical-shell subdomains that extend
to the outer boundary of the computational domain (located far 
from the holes).  The intermediate-zone cylindrical-shell and/or
rectangular-block subdomains must overlap both the inner and outer
spherical-shell subdomains in these grid structures to cover the
computational domain completely.  These grid structures are 
quite efficient, and have allowed our group
to perform long stable inspiral
calculations for a variety of simple binary black-hole
systems~\cite{Scheel2006, Pfeiffer-Brown-etal:2007,
  Boyle2007,Buonanno:2009qa}, and also simulations of the merger and
ringdown phases of a few of these simple
cases~\cite{Scheel2008,Chu2009}.  The overlap regions in these grid
structures are well behaved in these successful cases.  But in
more generic inspiral and merger simulations, these overlap
regions become significant sources of numerical error and
instability. 

\begin{figure}
\centerline{\includegraphics[width=3in]{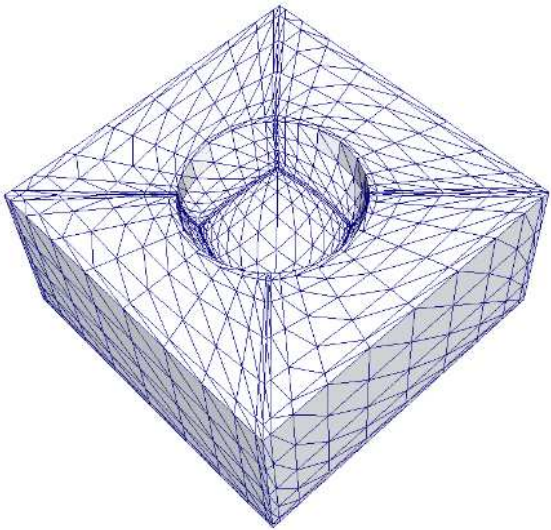}}
\caption{\label{f:CubeSphereB} Illustrates a basic element of the grid
  structure used to fill the 3D volume between a sphere and a
  concentric cube.}
\end{figure}

\begin{figure}[t]
\centerline{\includegraphics[width=3in]{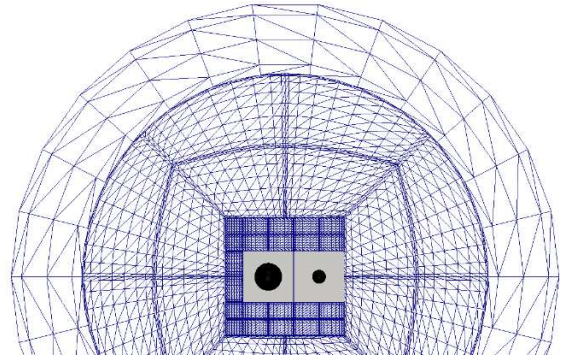}}
\caption{\label{f:SphereCGrid} Illustrates the intermediate grid
  structure surrounding the two black holes, consisting of a set of
  rectangular blocks placed around the cubic block containing each
  hole.  This collection of blocks filles up a cube which is
  surrounded by a series of cubed-sphere layers (two shown) 
  all surrounded
  by a series of spherical shells (one shell shown) that are centered
  on the center of mass of the binary system.}
\end{figure}

The overlap-related
instability discussed above is most likely caused by the method we
use to exchange information between adjacent subdomains.
We set the incoming characteristic fields, whose values are determined
by interpolation from the adjacent subdomain, using a penalty
method~\cite{Hesthaven1997, Hesthaven1999, Hesthaven2000,
  Gottlieb2001}.  This penalty method for imposing boundary conditions
was derived explicitly for the case where adjacent subdomain
boundaries touch but do not overlap.  One possibility for resolving
this instability would be to re-derive the appropriate penalty
boundary terms for the case of overlapping subdomains.  We have chosen
the simpler option of constructing new grid structures without
overlapping subdomains.  We do this by changing the intermediate-zone
grid structure from the cylindrical-shells and/or rectangular blocks
used previously into a series of deformed rectangular blocks whose
boundaries conform to the spherical shells used near the inner and
outer boundaries.  The new type of grid element needed for this
construction is illustrated in Fig.~\ref{f:CubeSphereB}, which shows
how the 3D volume between a sphere and a concentric cube can be mapped
into six deformed cubes.  This construction is based on the
``cubed-sphere'' coordinate representations of the sphere (i.e. the
mapping of the six faces of the cube onto the sphere as illustrated in
Fig.~\ref{f:CubeSphereB}), that is widely used in atmospheric and
geophysical modeling~\cite{Sadourny1972, Ronchi1996, Rancic1996,
  Taylor1997}.  Three-dimensional grid structures based on
cubed spheres have also been used in other types of simulations,
including a few black-hole simulations~\cite{Lehner:2005bz,
  Pazos:2009vb}.
\begin{figure}[t]
\centerline{\includegraphics[width=3in]{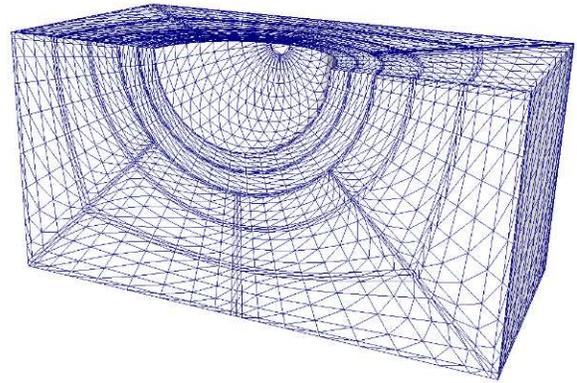}}
\caption{\label{f:SphereBGrid} Grid structure
  around each black hole.  A series of
  spherical shells (two shells shown) is surrounded by a series of 
  cubed-sphere layers
  (three shown) to fill up a cube.  This figure illustrates
  one quarter of the cube structure surrounding one of the black holes.}
\end{figure}
Our new grid structure is illustrated in Figs.~\ref{f:SphereBGrid} and
\ref{f:SphereCGrid}.  It consists of a series of spherical-shell
subdomains and several layers of cubed-sphere subdomains surrounding
each black hole, as illustrated in Fig.~\ref{f:SphereBGrid}.  The two
cubic blocks containing the black holes are surrounded by a series of
rectangular-block subdomains to fill out the volume of a large cube.
The center of this large cube is placed at the center-of-mass of the
binary system, as illustrated in Fig.~\ref{f:SphereCGrid} for a system
having a 2:1 mass ratio.  This large cube is connected to the outer
spherical-shell subdomains by several additional layers of
cubed-sphere subdomains, also illustrated in Fig.~\ref{f:SphereCGrid}.

Previous attempts to use this
type of cubed-sphere grid structure in SpEC
have always proven unsuccessful, due to numerical instabilities at the
interdomain boundaries.  These previous attempts 
 used no spectral
filtering, or a two-thirds anti-aliasing  filter, 
in the cubed-sphere subdomains.  It has been shown, however,
that an appropriate filter is required for stability (and improved
accuracy) of Chebyshev polynomial spectral expansions, such as those
used in these cubed-sphere subdomains.  Such a filter
must satisfy two important conditions.  The first criterion
is that when represented in physical space, this filter
must act like a dissipation term in the evolution equations that 
vanishes on the subdomain boundaries (to avoid conflicting with the
boundary conditions).  The second criterion is that it must set the 
highest spectral coefficient to zero. Adding this type of
spectral filtering turns out to be one of the key elements
in improving our numerical method enough to make possible the binary
black-hole simulations described in
Sec.~\ref{s:MergerRingdownSimulations}.  The needed spectral filter is
applied to each field $u(x,t)$ that is expanded as a sum of Chebyshev
polynomials:
\begin{eqnarray}
u(x,t)=\sum_k u_k(t)T_k(x).
\label{eq:ChebyshevExpansion}
\end{eqnarray}
These are used for the radial-coordinate expansions in the
spherical-shell subdomains, and for all three spectral-coordinate
expansions in the rectangular-block and the cubed-sphere subdomains.
Each spectral coefficient $u_k$ in these expansions is filtered
according to the expression,
\begin{eqnarray}
{\cal F}(u_k) = u_k e^{-\alpha (k/k_\mathrm{max})^{2p}},
\end{eqnarray}
after each time step.  This filter satisfies the first
necessary filter criterion by adding a
$2p^\mathrm{th}$-order dissipation term to the evolution equations
which vanishes on each subdomain boundary~\cite{Gottlieb2001}.  For
the binary black-hole simulations presented here, we use the filter
parameters $\alpha=36$ and $p=32$.  The choice $\alpha=36$ guarantees
that this filter satisfies the second necessary filter criterion
by ensuring that the highest spectral coefficient is set to
(double precision) zero, while the choice $p=32$ insures that
this is a very mild filter that only influences the largest few
spectral coefficients. This new filter is
applied wherever Chebyshev expansions, Eq.~(\ref{eq:ChebyshevExpansion}),
are used. 
We use the same filter we have used in previous binary black-hole
simulations in the angular directions in the spherical-shell subdomains: 
we set to zero the top four
$\ell$-coefficients in the tensor spherical-harmonic represention of
each dynamical field~\cite{Kidder2005,Scheel2006}.

\subsection{Adaptive Conforming Grid}
\label{s:AdaptiveConformingGrid}

One of the important developments that allowed 
our group (several years ago)
to begin performing successful binary black-hole inspiral simulations
was the introduction of the dual-coordinate-frame evolution
method~\cite{Scheel2006}.  This method uses two distinct coordinate
frames: One is a non-rotating and asymptotically Cartesian coordinate
system used to construct the tensor basis for the components of the
various dynamical fields evolved by our code.  The second is a
coordinate system chosen to follow (approximately) the motions of the
black holes.  We fix the computational grid to this second coordinate
frame, and solve the evolution equations for the ``inertial-frame''
tensor field components as functions of these ``grid-frame''
coordinates.  The map ${\cal M}$ connecting these grid frame
coordinates $\bar x^i$ to the inertial frame $x^i$, can be written
as the composition of more elementary maps:
\begin{eqnarray}
{\cal M} = {\cal M}_{\rm K} \circ {\cal M}_{\rm S}.
\label{eq:MapComposition}
\end{eqnarray}
${\cal M}_{\rm K}$ represents a ``kinematical'' map that keeps the
centers of the black holes located (approximately) at the centers of
the excised holes in our grid structures
(cf. Sec.~\ref{s:GridStructureFiltering}), and ${\cal M}_{\rm S}$ is a
``shape-control'' map that makes the grid conform (approximately) to
the shapes of the black holes.  The third major technical development
(which makes it possible now for us to perform robust merger
simulations) consists of improvements in the choice of the
shape-control map, ${\cal M}_{\rm S}$, and improvements in the way
this map is adapted to the dynamically changing shapes of the black
holes.

The kinematical map ${\cal M}_{\rm K}$ can itself be decomposed
into more elementary maps:
\begin{equation}
{\cal M}_{\rm K} = {\cal M}_{\rm T} \circ
            {\cal M}_{\rm E} \circ
            {\cal M}_{\rm R}.
\end{equation}
The maps ${\cal M}_{\rm T}$, ${\cal M}_{\rm E}$, and ${\cal M}_{\rm
  R}$ each move the centers of the black holes, but do not 
significantly distort
their shapes: The map ${\cal M}_{\rm T}$ translates the grid to
account for the motion of the center of the system
due to linear momentum being exchanged with the
near field~\cite{Lovelace:2009} and being emitted in gravitational
radiation. 
The map ${\cal M}_{\rm E}$ does a
conformal rescaling that keeps the coordinate distance between the
centers of the two black holes fixed in the grid frame, as they
inspiral in the inertial frame. And ${\cal M}_{\rm R}$ rotates
the frame so that the centers of the two black holes remain on the
$\bar x$ axis in the grid frame, as they move along their orbits in
the inertial frame.  These maps are also used during inspiral
simulations in the same way we use them for our merger
simulations.  So here we focus on the new shape-control map ${\cal
  M}_{\rm S}$, which is one of the critical new developments that made
the merger simulations reported in Sec.~\ref{s:MergerRingdownSimulations}
 possible.

The interiors of the black holes are excised in our simulations at the
spherical boundaries shown in Figs.~\ref{f:SphereBGrid} and
\ref{f:SphereCGrid}.  This is possible because, \emph{if this excision
  boundary is chosen wisely}, the spacetime inside this boundary can
not influence the spacetime region covered by our computational grid.
For hyperbolic evolution systems, like the generalized harmonic form
of the Einstein equations used in our code~\cite{Lindblom2006},
boundary conditions must be placed on each incoming characteristic
field at each boundary point.  Apparent horizons are surfaces that are
often used to study black holes, because they can be found numerically
in a fairly straightforward way, and because (if they exist at all)
they are always located within the true event horizons.  If the
excision bounaries for our computational domain were placed exactly on
the apparent horizons, then all the characteristic fields of the
Einstein system would be outflowing (with respect to the computational
domain) since apparent horizons are spacelike (or null) hypersurfaces.
Boundary conditions would not be needed on any field at such
boundaries.  Unfortunately we are not able to place excision
boundaries precisely on the apparent horizons.  So the best that can
be done numerically is to place them slightly inside the apparent
horizons (i.e., the apparent horizons must remain within the
computational domain), if we are to avoid the need for boundary
conditions on these excision surfaces.

However, if an excision boundary is placed {\em inside} an apparent
horizon, the outflow condition is no longer automatic or 
simple: the condition
depends on the shape and the location of the excision boundary, its
motion with respect to the horizon, and the gauge.  In our
simulations, the excision boundaries are kept somewhat inside the
apparent horizons, so the outflow condition can---and does---fail if
we are not careful.  One reason (and probably the principal reason) we
need to control the shapes of the computational domains through the
map ${\cal M}_{\rm S}$ is to keep pure outflow conditions on all the
dynamical fields at the excision boundaries.  We do this, as described
in more detail below, by requiring that the excision boundaries
closely track the shapes and sizes of the horizons.  

Another (probably secondary) reason that shape and size control are
needed in our numerical simulations is related to finite numerical
resolution.  If the excision boundaries had different shapes than the
horizons, then some points on the boundaries would be located deeper
into the black-hole inerior and hence closer to the spacetime
singularity.  Higher numerical resolution would be needed at these
points, and for fixed resolution, the amount of constraint violation
and errors in the solution would be largest there.  In many situations
we find that numerical instabilities at these points cause our
simulations to fail.  This mode of failure can be eliminated by
keeping the shapes and sizes of the boundaries close to those of the
horizons.

We have decomposed the map ${\cal M}$, which connects the grid-frame
coordinates $\bar x^i$ with inertial-frame coordinates $x^i$, into
kinematical and shape-control parts: ${\cal M}={\cal M}_{\rm K}\circ
{\cal M}_{\rm S}$.  It will be convenient to have a name for the
intermediate coordinate system whose existence is implied by this
split. The map ${\cal M}_{\rm K}$ connects the inertial-frame
coordinates $x^i$ with coordinates $\tilde x^i$ in which the
centers of the black holes are at rest (approximately). 
So it is natural to call these intermediate-frame coordinates, 
$\tilde x^i$, the ``rest frame'' coordinate system.  
The shape control map ${\cal M}_{\rm S}$ connects the grid-frame 
coordinates $\bar x^i$ with
these rest-frame coordinates $\tilde x^i$.

It is useful to express the shape-control map ${\cal M}_{\rm S}$ as
the composition of maps acting on each black hole individually: ${\cal
  M}_{\rm S}={\cal M}_{\rm S_1}\circ {\cal M}_{\rm S_2}$.  Such
individual black-hole shape-control maps can be written quite
generally in the form
\begin{eqnarray}
\label{eq:DistortionMap1}
\tilde{\theta}_A &=&\bar \theta_A,\\
\label{eq:DistortionMap2}
\tilde{\phi}_A   &=&\bar \phi_A,\\
\label{eq:DistortionMap3}
\tilde{r}_A      &=& \bar r_A - f_A(\bar r_A,\bar \theta_A,\bar \phi_A)
                \nonumber \\
                &&\qquad\qquad\times\sum_{\ell=0}^{\ell_{\rm max}}
                \sum_{m=-\ell}^{\ell} \lambda^{\ell m}_A(t)
                Y_{\ell m}(\bar \theta_A,\bar\phi_A),\qquad
\end{eqnarray}
where $(\bar r_A,\bar \theta_A,\bar \phi_A)$ and
$(\tilde{r}_A,\tilde{\theta}_A,\tilde{\phi}_A)$ are, respectively, the
grid-frame and rest-frame
spherical polar coordinates centered at
the (fixed) grid-coordinate location of black hole ${\scriptstyle
  A}=\{1,2\}$.  The functions $f_A(\bar r_A,\bar \theta_A,\bar
\phi_A)$ are fixed functions of space in the grid frame.  We note that
these maps become the identity whenever $f_A(\bar r_A,\bar
\theta_A,\bar\phi_A)=0$, so making $f_A(\bar r_A,\bar
\theta_A,\bar\phi_A)$ vanish as $\bar r_A\rightarrow \infty$ ensures
that the distortion is limited to the neighborhood of each black hole.
The parameters $\lambda^{\ell m}_A(t)$ specify
the angular structure
of the distortion map.  They are determined by a feedback control
system (discussed in Appendix~\ref{s:control-systems})
that dynamically adjusts the
shape and overall size of the grid frame coordinates relative to the
rest-frame coordinates.

The Caltech/Cornell collaboration has used shape-control maps of this form,
Eqs.~(\ref{eq:DistortionMap1})--(\ref{eq:DistortionMap3}), in all 
of our recent binary black-hole simulations~\cite{Scheel2008,
Lovelace:2009,Chu2009}. 
In those cases the functions $f_A(\bar
r_A,\bar \theta_A,\bar \phi_A)$ were taken to be smooth functions of
$\bar r_A$ alone: roughly constant near each black hole and falling
off to zero rapidly enough that ${\cal M}_{{\rm S_1}}$ and ${\cal
  M}_{{\rm S_2}}$ approximately commute.  These functions had large
gradients which made it difficult to control the grid distortion near
the horizons without introducing additional unwanted distortions
elsewhere.  As a result, we were never able to achieve very robust
shape control using these maps.

One of our major breakthroughs, leading to the successful merger
results reported here, is an improvement in our choice of the map
functions $f_A(\bar r_A,\bar \theta_A,\bar \phi_A)$.  We now use
simpler functions $f_A(\bar r_A,\bar \theta_A,\bar \phi_A)$ 
that have much smaller gradients
and that are exactly zero outside (non-intersecting) compact regions 
surrounding each black hole.  This means the new maps ${\cal M}_{{\rm
    S_1}}$ and ${\cal M}_{{\rm S_2}}$ commute, exactly.  These
new maps can be defined everywhere between the
black holes because of the new non-overlapping
grid structure discussed in
Sec.~\ref{s:GridStructureFiltering}, but these improvements
are achieved at the expense
of smoothness: the new $f_A(\bar r_A,\bar \theta_A,\bar \phi_A)$ are
smooth except at subdomain boundaries where they may only be
continuous, not differentiable
\footnote{The idea of introducing non-smooth time-dependent mappings
was also suggested by Larry Kidder.}.  Fortunately, this lack of smoothness
at subdomain boundaries is not a problem for our basic evolution method. 
In our multidomain code the
equations are solved independently on each subdomain, and subdomains
communicate only by equating the appropriate characteristic fields at
their mutual boundaries.  These boundary conditions require no
differentiation, so the grid coordinates themselves need not be smooth
across these boundaries.

\begin{figure}
\includegraphics[width=0.25\textwidth]{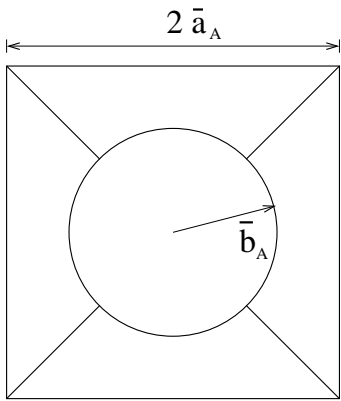}
\caption{\label{fig:DistortionMapGrid}
Two-dimensional illustration of the 
grid-frame coordinate domain on which the function 
$f_A(\bar r_A,\bar \theta_A,\bar\phi_A)$ is
defined in Eq.~(\ref{eq:DistortionMapF}).}
\end{figure}

The new $f_A(\bar r_A,\bar \theta_A,\bar \phi_A)$ are chosen to be
unity inside a sphere of radius $\bar b_A$ centered on black hole
$\scriptstyle A=\{1,2\}$, then decrease linearly with radius along a
ray through the center of the black hole, and finally vanish on the
surface of a centered cube of size $2 \bar a_A$.  A two-dimensional
sketch of this grid-frame domain is shown in
Fig.~\ref{fig:DistortionMapGrid}, which is merely an abstract version
of the type of grid structures we use around each black hole as
illustrated in Figs.~\ref{f:CubeSphereB} and \ref{f:SphereBGrid}.
This function $f_A(\bar r_A,\bar \theta_A,\bar \phi_A)$ can be
expressed analytically as
\begin{equation}
\label{eq:DistortionMapF}
f_A(\bar r_A, \bar\theta_A,\bar\phi_A) = \left\{
\begin{array}{cl}
1,& \quad\bar r_A \le \bar b_A, \\
\displaystyle
\frac{\bar r_A - \bar \rho_{A}}{\bar b_A - \bar \rho_{A}},
& \quad\bar b_A \le \bar r_A \le \rho_{A}, \\
0, & \quad\bar\rho_A \le \bar r_{A},
\end{array}
\right.
\end{equation}
where
\begin{eqnarray}
\label{eq:DistortionMaprmax}
&&\!\!\!\!\!\!\!
\bar \rho_{A}(\bar \theta_A,\bar \phi_A) = \bar a_A\\
&&\times\left[\mathrm{max}\left(|\sin\bar\theta_A\cos\bar\phi_A|,
                                    |\sin\bar\theta_A\sin\bar\phi_A|,
                                    |\cos\bar\theta_A|
                                    \right)\right]^{-1}\nonumber
\end{eqnarray}
is the value of $\bar r_A$ on the surface of the cube 
of size $2 \bar a_A$ centered on
black hole $\scriptstyle A = \{1,2\}$.  The
  different cases of the maxima which appear in the denominator of
  Eq.~(\ref{eq:DistortionMaprmax}) correspond to the different
  faces of this cube.  

We note that the choice of $f_A(\bar r_A,\bar \theta_A,\bar \phi_A)$
in Eq.~(\ref{eq:DistortionMapF}) implies that the shape-control map
${\cal M}_{\rm S}$, and hence the map between grid-frame and 
rest-frame coordinates is not smooth.  Since the kinematical map ${\cal
  M}_{\rm K}$ is smooth, this also imples that the full map ${\cal M}$
relating grid-frame to inertial-frame coordinates is not smooth
either.  This non-smoothness does not cause a problem for our basic
evolution method, but it implies that calculations 
that require smoothness must  
not be performed in the grid frame.
For example, the representation of a smooth apparent
horizon in grid coordinates will not be smooth.  
So apparent horizon finding and other
calculations that require smoothness are done
in the smooth rest-frame or inertial-frame
coordinates.

\begin{figure}
\centerline{\includegraphics[width=2.7in]{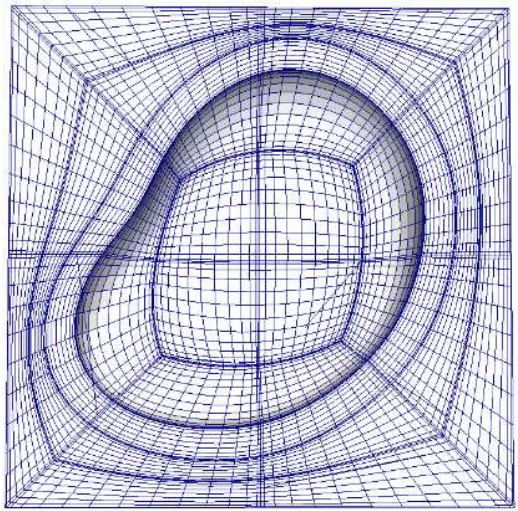}}
\caption{\label{f:InertialGrid} Illustrates the inertial frame
  representation of the grid structure around one black hole, just at
  the time of merger. This grid structure has been distorted in
  relation to the inertial frame by the shape-control 
  maps ${\cal M}_{\rm S}$ described here.}
\end{figure}

The angular structures of the shape-control maps defined in
Eq.~(\ref{eq:DistortionMap1})--(\ref{eq:DistortionMap3}) are
determined by the parameters $\lambda^{\ell m}_A(t)$.  These are
chosen dynamically to ensure that the shapes of the excision
boundaries match the evolving shapes of the apparent horizons.  These
maps also control the sizes of the excision boundaries by adjusting
$\lambda^{00}_A$, which are chosn to ensure that the excision
boundaries remain close to but safely inside the apparent horizons.
Figure~\ref{f:InertialGrid} illustrates the effect of this
distortion map on the structure of the grid surrounding
a black hole in one of the generic  merger simulations described in
Sec.~\ref{s:MergerRingdownSimulations}.  The choices of suitable
target shapes and sizes for these shape-control maps, and the feedback
control system that we use to implement these choices dynamically in
our simulations, are also among the critical technical advances that allow
us now to perform robust merger and ringdown simulations.  The details
of exactly how this is done are somewhat complicated, however, 
so we defer their
discussion to the Appendix~\ref{s:control-systems}.

\section{Merger and Ringdown Simulations}
\label{s:MergerRingdownSimulations}

In this section we present simulations of the merger and ringdown
of binary black-hole systems using our Spectral Einstein
Code (SpEC).
In Section~\ref{s:initial-data} we describe the binary black-hole initial
data sets that we evolve.  In Section~\ref{sec:evolutionprocedure} we
discuss our evolution algorithm, 
pointing out at which stage we employ the various
improvements described in Section~\ref{s:TechnicalDevelopments} and what
goes wrong when these improvements are not used.  Finally,
in Section~\ref{sec:results} we describe six different binary black-hole 
merger and ringdown simulations peformed using
these new methods. We present
snapshots of 
the shapes and locations of the horizons at various times
during the merger and ringdown, and
a demonstration of convergence of the constraint violations.

\subsection{Initial Data}
\label{s:initial-data}

Our initial data are in
quasi-equilibrium~\cite{Cook2002,Cook2004,Caudill-etal:2006} (see
also~\cite{Gourgoulhon2001,Grandclement2002}), and are built using the
conformal thin sandwich formalism~\cite{York1999,Pfeiffer2003b} with
the simplifying choices of conformal flatness and maximal slicing.
Quasi-equilibrium boundary conditions are imposed on spherical
excision boundaries for each black hole, with the lapse boundary
condition given by Eq.~(33a) of Ref.~\cite{Caudill-etal:2006}.  The
spins of the black holes are determined by boundary conditions on the
shift vector at each excision surface~\cite{Cook2004}.

This formalism for constructing initial data
also requires the
initial radial velocity $v_r$ of each black hole toward its partner
and an initial orbital angular frequency $\Omega_0$ (chosen to be
about the $z$ axis without loss of generality).  These parameters determine the
orbital eccentricity of the binary, and can be tuned by an iterative
method~\cite{Pfeiffer-Brown-etal:2007,Chu2009} to produce data with
very low eccentricity.  This has been done for most of the initial
data sets described here (cases A through D in Table~\ref{tab:Runs}), 
but is unnecessary for the purposes of the
present paper, since our goal here is simply to document our improved
methods for evolving binaries through merger.  For cases E and F,
$v_r$ and $\Omega_0$ are chosen roughly by using 
post-Newtonian formulae that ignore spins.
\begin{table}
\begin{tabular}{|c|cccc|}
\hline
Case & $M_2/M_1$ & $\vec S_1/M_1^2$ & $\vec S_2/M_2^2$ & $N_{\rm orbits}$ \\
\hline
A &1&0                              &0                                &16 \\
B &2&0                              &0                                &15 \\
C &1&$-0.4\hat{z}$                  &$-0.4\hat{z}$                    &11 \\
D &1&$\phantom{-}0.4\hat{z}$        &$\phantom{-}0.4\hat{z}$          &15 \\
E &2&$0.2(\hat{z}-\hat{x})/\sqrt{2}$&$-0.4(\hat{z}+\hat{y})/\sqrt{2}$ &8.5\\
F &2&$0.2(\hat{z}-\hat{x})/\sqrt{2}$&$-0.4(\hat{z}+\hat{y})/\sqrt{2}$ &1.5\\
\hline
\end{tabular}
\caption{\label{tab:Runs}
Black hole binary configurations run through merger and ringdown using the
methods presented here. For the initial spin parameters $\vec S_1/M_1^2$ 
and $\vec S_2/M_2^2$, the $\hat{z}$ direction is parallel to the orbital
angular momentum.
}
\end{table}

Table~\ref{tab:Runs} shows the mass ratio and initial spins for the
black hole binary configurations we evolve here.  The initial data and
first 15 orbits of inspiral for case A are identical to those
presented in~\cite{Boyle2007,Scheel2008}.  The initial data and first
9 orbits of inspiral for case C are identical to those presented
in~\cite{Chu2009}. Cases E and F are fully generic, with unequal masses and
unaligned spins, and exhibit precession and radiation-reaction 
recoil.

\subsection{Evolution procedure}
\label{sec:evolutionprocedure}

In this section we summarize our procedure for evolving black hole binary
systems through merger and ringdown, concentrating on the improvements
discussed in Section~\ref{s:TechnicalDevelopments}.   This procedure can
be divided into several stages, 
beginning with the early inspiral and extending
through ringdown:
\begin{enumerate}
\item Evolve early inspiral using ``quasi-equilibrium'' gauge.
\item Transition smoothly to damped harmonic gauge. \label{item:dampedharmonic}
\item Eliminate overlapping subdomains. \label{item:nooverlap}
\item Turn on shape control.
\item Replace remaining inner spherical shells with cubed spheres.
\item Turn on size control just before holes merge.
\item After merger, interpolate to a single-hole grid to run through ringdown.
\end{enumerate}
The above ordering of these stages is not the only possible choice: 
we have exchanged the order of
some of these stages without trouble.
For example, for the runs
described here we perform stage~\ref{item:nooverlap} at the instant we
start stage~\ref{item:dampedharmonic}.
The transitions between these stages are relatively simple, require
little or no fine tuning, and can be automated. 
We now discuss each of these stages in turn.  For several of these
stages, we will illustrate the effects of new improvements 
(Section~\ref{s:TechnicalDevelopments}) for
one particular evolution, case F of Table~\ref{tab:Runs}.

\subsubsection{Early inspiral}
\label{sec:inspiral}

The gauge is chosen early in the evolution following
the procedure of Ref.~\cite{Boyle2007}:  the initial data
are constructed in quasi-equilibrium, so we choose the initial $H_a$ 
to make the time derivatives of the lapse and shift 
zero.  We attempt to maintain this quasi-equilibrium condition by demanding that
$H_a$ remains constant in time in the grid frame in the following sense:
we require
\begin{equation}
\label{eq:QuasiEquilibriumGauge}
\partial_{\bar{t}} \tilde{H}_{\bar{a}} = 0,
\end{equation}
where $\tilde{H}_a$ is a tensor (note that
$H_a$ is not a tensor) defined so that $\tilde{H}_a=H_a$ in
the inertial frame.  The bars in Eq.~(\ref{eq:QuasiEquilibriumGauge}) 
refer to the grid frame coordinates.
We refer to this as ``quasi-equilibrium'' gauge, although strictly speaking
this gauge maintains quasi-equilibrium only 
when the spin directions are constant in the grid frame.

For cases A--D, the simulations follow many orbits using this
quasi-equilibrium gauge.  For cases E and F, in which the spins 
 change direction in the grid frame,
this gauge is not appropriate and causes difficulties, 
so for these cases we transition to
harmonic gauge $H_a=0$ very early in the inspiral.

\begin{figure}
\centerline{\includegraphics[height=3in]{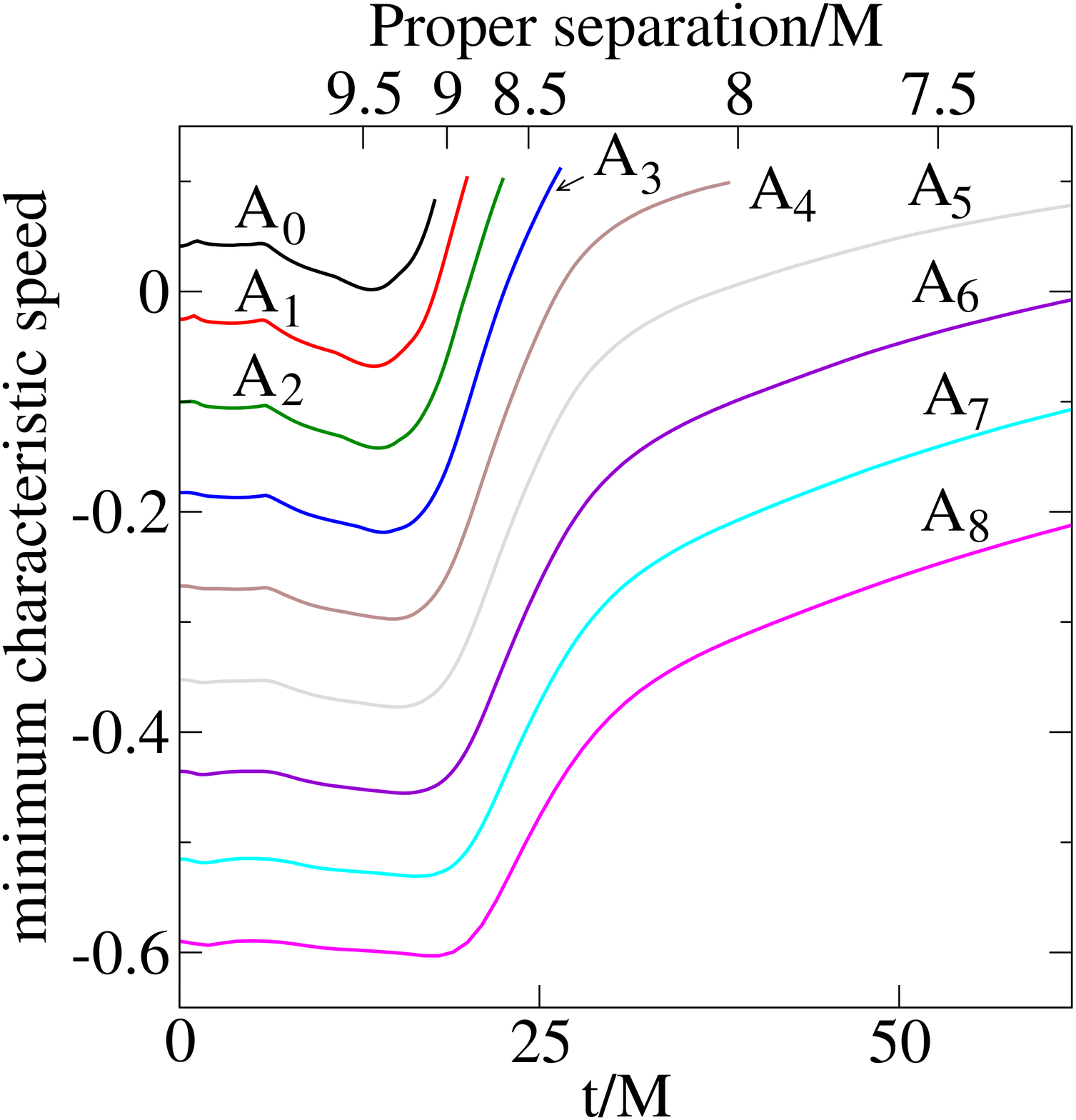}}
\caption{\label{f:CharSpeedsAhA} 
  The minimum of the outgoing (into
  the hole) characteristic
  speeds on the nine innermost spherical grid boundaries (labeled
  $A_0 \ldots A_8$) around the larger
  black hole, for part of run F.  
 }
\end{figure}

During inspiral, the map ${\cal M}_{\rm E}$ uniformly contracts
the grid so that the grid-coordinate distance between the centers of the
two black holes remains constant.  Because this contraction is uniform,
the spherical-shell subdomains inside each apparent horizon shrink relative
to the horizon.  This means that the excision boundary 
(the inner boundary of the innermost spherical shell) inside each
black hole moves further into the strong-field region in the interior,
towards the singularity.  If this motion continues unchecked, 
gradients on the innermost spherical shell 
increase until the solution
is no longer resolved.  However, the inner
boundary of the {\em next-to-innermost} 
spherical shell is also shrinking, and it
eventually shrinks enough that the characteristic fields on this boundary 
all become outflowing (into the black hole); when this happens, the innermost
spherical shell no longer influences the exterior solution, so we
simply drop that shell from the domain.  
This shell-dropping process occurs automatically as successive spherical
shells shrink relative to the horizons, 
This process is illustrated in Figure~\ref{f:CharSpeedsAhA},
which shows the minimum of the (into the hole) characteristic
speeds on spherical grid boundaries around one of the
black holes.  At any given time, the innermost of these boundaries
is an excision boundary.
All characteristic speeds at all points
on the excision boundary (and hence the
minimum) must be positive for the excision algorithm
to be well-posed. The innermost shell is dropped whenever the characteristic
speeds are positive on the inner boundary of
the next shell. For example, surface $A_3$ in Figure~\ref{f:CharSpeedsAhA}
is the excision boundary at $t=24M$, 
but after the characteristic speeds on $A_4$ become positive (around $t=26M$),
the innermost shell is dropped and $A_4$ becomes the excision boundary.
Similarly, $A_5$ becomes the excision boundary at $t=38$, and so on.
It is important that the characteristic speed on
surface $A_n$ in Figure~\ref{f:CharSpeedsAhA} does not become negative 
before the speed on surface $A_{n+1}$ becomes positive;
otherwise this shell-dropping procedure will fail.
Shell dropping works well during  most of the evolution
but it fails in the last stage
before merger, which is discussed in Section~\ref{sec:sizecontrol}.

When describing mergers, it is useful to introduce a measure
of distance between the black holes. 
One such measure of distance is indicated
on the horizontal axis at the top of 
Figure~\ref{f:CharSpeedsAhA}.
This is the spatial (in each time slice) 
proper separation between the apparent horizon surfaces, obtained
by integrating along the rest-frame $\tilde{x}$ axis (recall that in
the rest frame, the centers of the horizons always remain along this axis).  
This is not the true
proper separation between the horizons because we do not minimize over
all possible paths between all possible pairs of points on the two surfaces;
however, we use this only as an approximate
measure of how far the black
holes are from merger.  Note that the horizons touch each other when
(or possibly before) this proper separation measure falls to zero.

It turns out that the improvements discussed in
Section~\ref{s:TechnicalDevelopments} are unnecessary during
early inspiral for cases A--D
and are not yet used at this stage.
For instance, the domain decomposition uses overlapping
cylinders and spherical shells, and not the cubed-sphere subdomains
described in Sec.~\ref{s:GridStructureFiltering}.  We find no problems
while the holes are far enough apart that they remain roughly in equilibrium.
However, the binary eventually becomes extremely dynamical, and the black holes 
become significantly distorted.  Unless the algorithm is modified,
the simulation fails
(typically because of large constraint violations) before the black holes
merge.

Typically, it is possible to extend the quasi-equilibrium inspiral
stage until the proper separation 
 decreases to about $7M$ or $8M$,
but the next stages, Secs.~\ref{sec:eliminate-overlapping-subdomains} 
and~\ref{sec:harmonic-gauge},
can be done sooner if desired. (For example, cases E and
F use a non-overlapping cubed-sphere grid from $t=0$.) 
For the simulations presented here, the length of this
quasi-equilibrium inspiral stage is variable.
For instance, for case A in Table~\ref{tab:Runs} this stage lasts
for 15 orbits, but for case F we transition to damped harmonic gauge
and eliminate overlapping subdomains
starting at $t=0$ because the black holes are initially very
close together.

\subsubsection{Transition to damped harmonic gauge}
\label{sec:harmonic-gauge}

When the inspirling black holes reach a proper separation of about 
$8 M$, we smoothly
transition from
quasi-equilibrium gauge to damped harmonic gauge 
at $t=t_g$ ($g$ stands for ``gauge'')
by choosing
\begin{eqnarray}
H_{a}(t) &=& \tilde{H}_{a}(t) e^{-(t-t_g)^4/\sigma_g^4}
\nonumber \\ &&
+ \mu_L\log\left(\frac{\sqrt{g}}{N}\right)t_a-\mu_S N^{-1}g_{ai}N^i,
\label{eq:QuasiEquilibriumGaugeFalloff}
\end{eqnarray}
where $\tilde{H}_{a}(t)$ is the value of $H_a(t)$ obtained
from the quasiequilibrium gauge condition~(\ref{eq:QuasiEquilibriumGauge}), 
and $\sigma_g$ is a constant.
The last line in Eq.~(\ref{eq:QuasiEquilibriumGaugeFalloff}) is the damped
harmonic gauge condition, Eq.~(\ref{e:FullDampedWaveGauge}).
The values of $\mu_L$ and $\mu_S$ are set according to 
Eq.~(\ref{eq:GaugeDampingFactor}), using
\begin{equation}
\label{eq:GaugeDampingFactorRollon}
\mu_0 = \left\{
\begin{array}{cl}
0, & t<t_d \cr
1-e^{-(t-t_d)^2/\sigma_d^2}, & t>t_d
\end{array}\right.,
\end{equation}
where $t_d$ and $\sigma_d$ are constants.
It is necessary to choose $\sigma_g$ and $\sigma_d$ sufficiently large or else
the gauge becomes unnecessarily dynamical, possibly resulting in incoming
characteristic fields on the excision boundary.

For all runs shown here, $t_d$ corresponds approximately
to a proper separation of $8M$. We
use the values $\sigma_g = 20M, \sigma_d=50M$
for the equal mass binaries (cases A, C, and D),
$\sigma_g=15M,\sigma_d=100M$ for case B,
and $\sigma_g=25M,\sigma_d=50M$ for case F.
For these cases, $t_g=t_d$.
For case E, the gauge is rolled off to harmonic early
in the inspiral, so at $t=t_d$ we need only to roll on
the damped harmonic gauge, and we use $\sigma_d = 40M$.
The constants in Eqs.~(\ref{eq:QuasiEquilibriumGaugeFalloff})
and~(\ref{eq:GaugeDampingFactorRollon}) 
can be chosen quite flexibly: the
only constraint is that $\sigma_g$ and $\sigma_d$ must be
large enough to avoid significant spurious gauge dynamics, but small enough
so that the simulation is using damped harmonic gauge before the black holes
approach each other too closely. 
The  $\sigma_g$ and $\sigma_d$ can always be made longer by choosing
earlier transition times $t_g$ and $t_d$.

\subsubsection{Eliminate overlapping subdomains}
\label{sec:eliminate-overlapping-subdomains}

Although the domain decomposition consisting of overlapping
spherical shells, cylinders, and cylindrical shells is quite efficient
for the early inspiral, it often suffers from numerical instabilities,
particularly as the black holes approach merger.
Therefore, we regrid onto the cubed-sphere subdomains described in
Sec.~\ref{s:GridStructureFiltering}.  Use of these non-overlapping
subdomains eliminates these remaining instabilities, and effectively 
increases our resolution.  For most of
the runs described here, we happen to regrid
at the same time $t_d$ that we begin the transition to damped
harmonic gauge.  However, for runs E and F, 
the entire simulation uses the non-overlapping
cube-sphere subdomains.
As far as we know, there is no reason that regridding
cannot occur at times other than $t_d$ or $t_g$.

\subsubsection{Shape control}
\label{sec:turn-on-shape-control}

The next stage is to turn on shape control by
introducing the map ${\cal M}_{\rm S}$ 
defined in Eqs.~(\ref{eq:DistortionMap1}--\ref{eq:DistortionMaprmax}).
The parameters $\lambda_A^{\ell m}$ that appear in this map
are determined (via a feedback control system) by
Eqs.~(\ref{eq:lambda-def}) and~(\ref{eq:ShapeControlBigLambdalm})
for $\ell>0$, and $\lambda_A^{00}$ is set to zero.
This map deforms the grid 
so that the boundaries of the spherical shells inside the horizons, including
the excision boundaries, are mapped to closely match the horizons' shapes.  

We do not turn on shape control until time 
$t_g + 2.5 \sigma_g$, 
which is when the coefficient multiplying the gauge term $\tilde{H}_{a}(t)$
in Eq.~(\ref{eq:QuasiEquilibriumGaugeFalloff}) becomes smaller than 
double-precision
roundoff.  This ensures that $H_a$ remains a smooth function of
space in inertial coordinates.   If shape control were turned on too
early, $H_a$ would fail to be everywhere smooth
because the quasi-equilibrium-gauge quantity
$\tilde{H}_{\bar{a}}$ 
is a smooth function of space in the {\em grid-frame} 
coordinates (and its numerical value is constant in time for
each grid point),
and the deformation is only $C^0$ across subdomain boundaries.

\begin{figure}
\centerline{\includegraphics[height=3in]{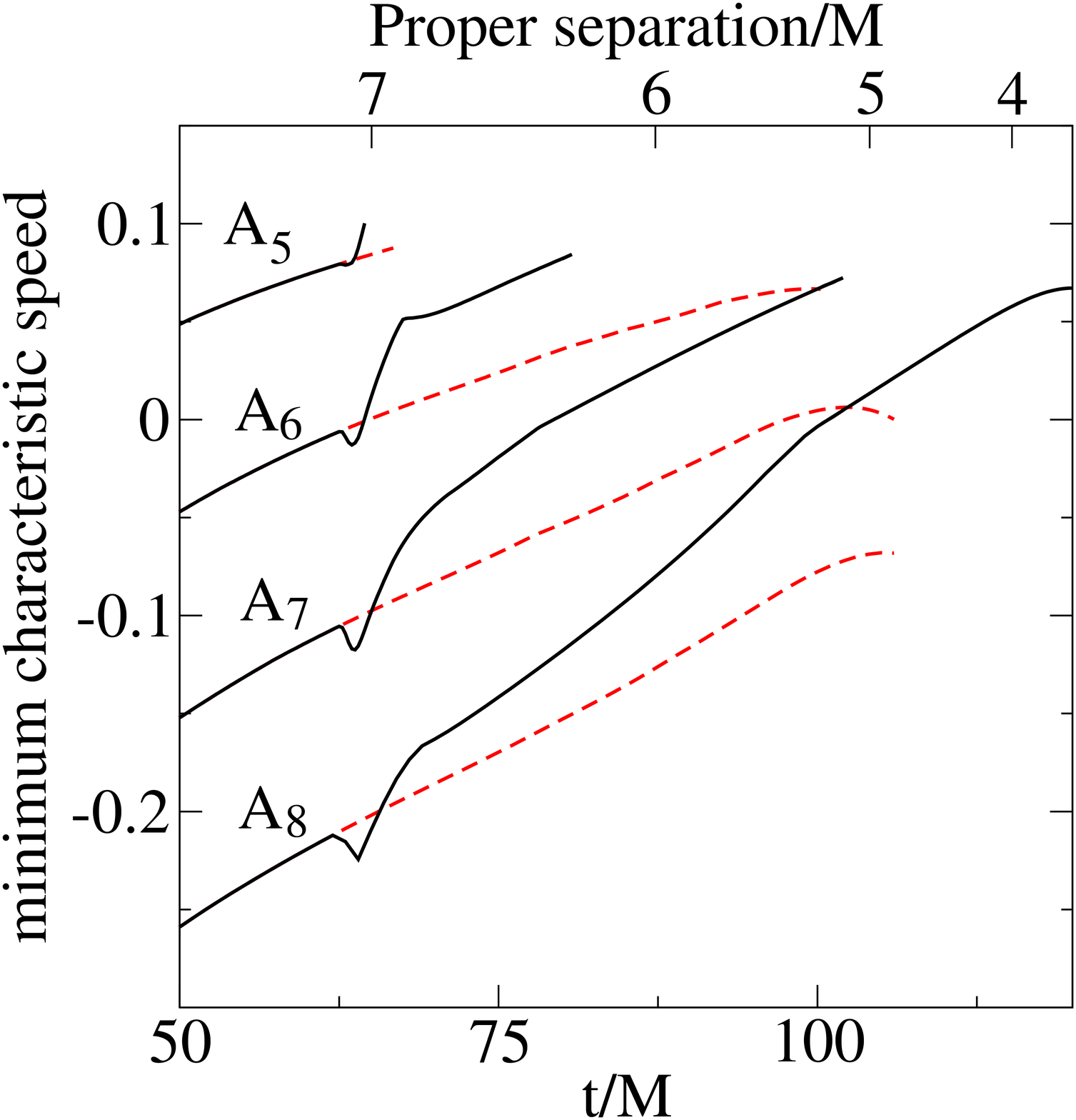}}
\caption{\label{f:CharShapeControl} The minimum of the outgoing (into
  the hole) characteristic
  speeds on the remaining four
  spherical-shell boundaries (labeled $A_5 \ldots A_8$) around the larger
  black hole, for a portion of run F.  
  The dashed curves correspond to runs without
  shape control, and the solid curves correspond to runs with shape
  control turned on at $t=62M$.
  The run without
  shape control terminates at proper separation $5.2 M$ (time $t=106M$)
  because the excision algorithm fails on surface $A_7$.
 }
\end{figure}

Shape control is necessary for the shell-dropping procedure described in
Section~\ref{sec:inspiral} to remain successful 
as the holes approach each other.
For a shell to be dropped, the 
{\em entire} inner boundary of that shell must remain an outflow surface
until the {\em entire} inner boundary of the next shell becomes an outflow
surface.
If these boundaries
do not have the same shape as the horizon, then typically
some portion of either boundary moves too close or too far from the horizon,
and violates  this condition.
An example is shown in Figure~\ref{f:CharShapeControl}.
Like Figure~\ref{f:CharSpeedsAhA}, Figure~\ref{f:CharShapeControl}
shows the minimum characteristic
speeds on the boundaries of inner spherical shells for a portion
of case F, but Figure~\ref{f:CharShapeControl} shows two separate evolutions: 
one with shape control turned on at $t=62M$ (solid curves)
and one with no shape control (dashed curves).
Without shape control, the minimum characteristic speed on surface
$A_7$ stops
growing around $t=100M$ and becomes negative soon thereafter.  Because
$A_7$ is the excision surface at $t\sim 100M$, the excision algorithm
becomes ill-posed and the simulation stops at $t=106M$, well before
merger (which occurs at $t\sim 128M$).  In the simulation with active
shape control, the same boundary $A_7$ shows no such problems---in
fact it is dropped at $t=102M$ when all speeds on the next shell,
$A_8$ become positive.

\subsubsection{Eliminate inner spherical shells}
\label{sec:nomorespheres}

\begin{figure}
\centerline{\includegraphics[height=3in]{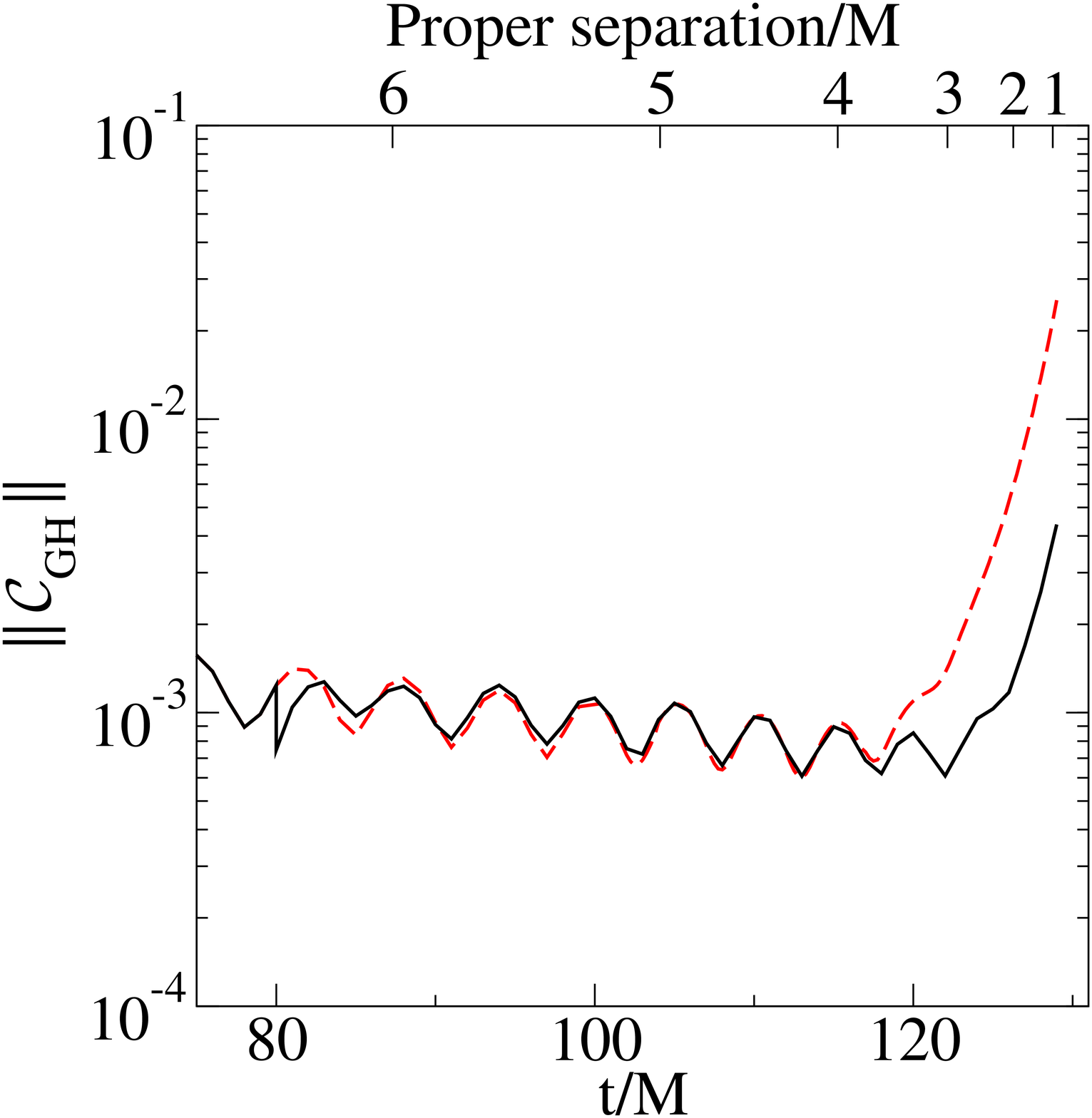}}
\caption{\label{f:NoCubeSpheres} 
  Constraint norm $||{\cal C}_{GH}||$ as a function of time
  for a portion of case F.  The dashed curve corresponds to a run with
  spherical shells around each hole, and the solid curve corresponds to
  an identical run with these inner spherical shells replaced by
  cubed-sphere shells at $t=80M$. 
  The constraint norm grows larger and 
  begins to grow earlier when spherical shells are used.
 }
\end{figure}

Spherical-shell subdomains are extremely efficient when the solution is
nearly spherically symmetric, because these subdomains use $Y_{lm}$ basis
functions, which are well-suited for nearly spherical functions.
However, once the holes become
sufficiently distorted, it is difficult to resolve the region near each
horizon using spherical shells.  At this point in the evolution, we
replace the inner spherical shells with cubed-sphere shells. 
Figure~\ref{f:NoCubeSpheres} shows the constraints as a function of time for
two evolutions of case F: 
one with inner spherical shells (dashed curves), and another 
in which the inner spherical shells were replaced
with cubed-sphere shells at $t=80M$ (solid curves).
The quantity plotted is $||{\cal C}_{GH}||$, the 
$L^2$ norm over all the constraint fields of our first-order
generalized harmonic system, normalized by the $L^2$ norm over the
spatial gradients of all dynamical fields (see Eq. (71) of
Ref.\cite{Lindblom2006}).  The $L^2$ norms are taken over the portion
of the computational volume that lies outside the apparent horizons.
For the spherical-shell case in Figure~\ref{f:NoCubeSpheres}, the
constraints in the spherical shells grow large enough that the simulation
terminates at $t=129.5M$, when the proper separation has fallen
to $0.5 M$.  Note that these simulations
typically proceed
through merger even if we do not replace spherical
shells with cubed-sphere shells. So while
this replacement step is not strictly
necessary, 
eliminating spherical shells improves the accuracy
of the simulation (as measured by the constraint quantity
$||{\cal C}_{GH}||$) by almost an order of magnitude.

\subsubsection{Size control}
\label{sec:sizecontrol}

\begin{figure}
\centerline{\includegraphics[height=3in]{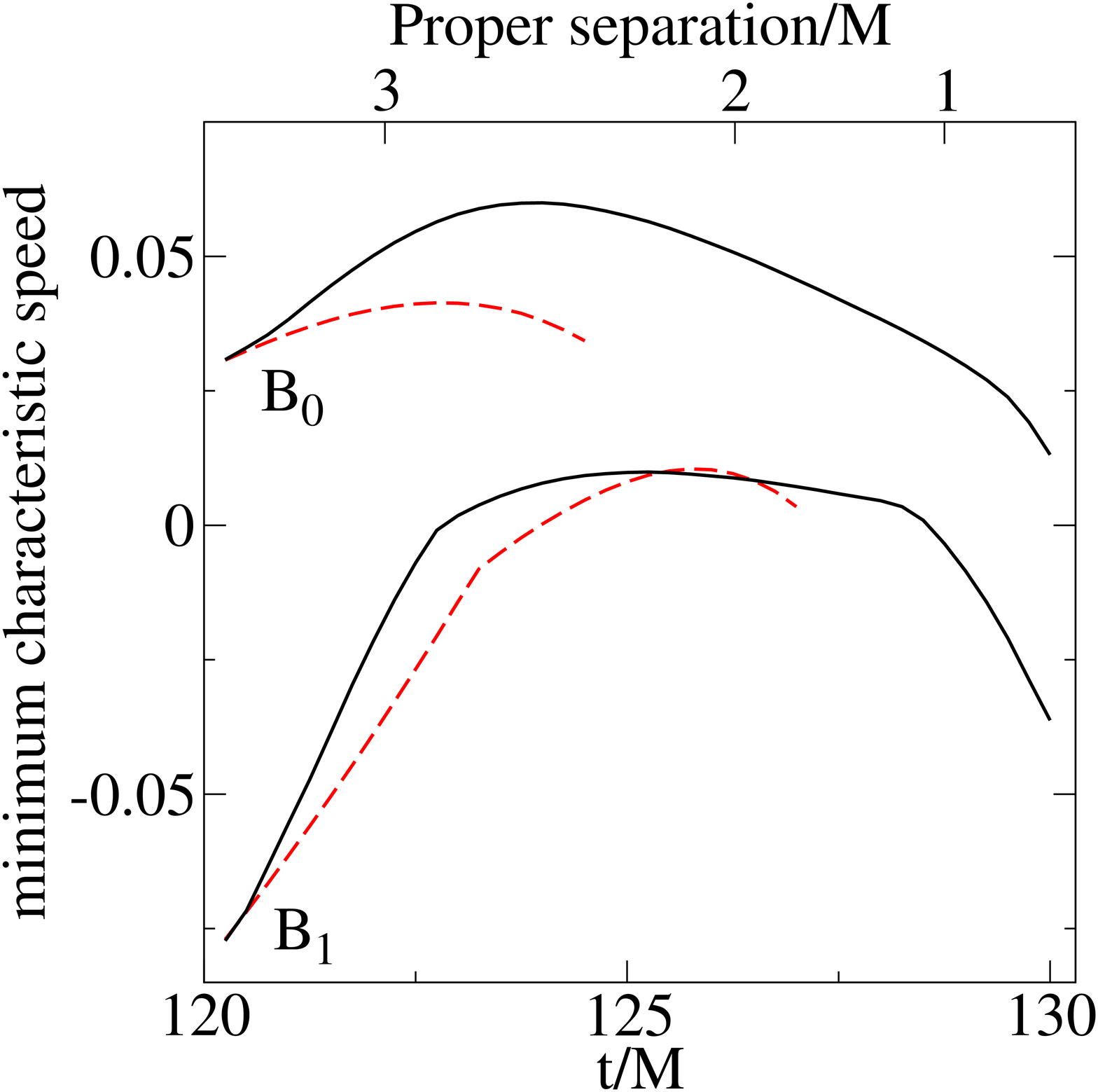}}
\caption{\label{f:CharSizeControl} The minimum of the outgoing (into
  the hole) characteristic
  speeds on the two innermost cubed-sphere boundaries (labeled
  $B_0$ and $B_1$) around the smaller black hole, for a portion of run F.  
  The dashed curves correspond to runs without
  size control, and the solid curves correspond to runs with size
  control turned on at $t=120.2M$.  
  Without size control, the innermost cubed-sphere is dropped
  at $t=124.5M$ and $B_1$ becomes the excision boundary. However,
  the characteristic speeds become negative on $B_1$ at $t=127M$
  and excision fails.  With size control, the characteristic speeds
  on the excision boundary $B_0$ remain positive until
  proper separation $0.075M$,  well after merger.
 }
\end{figure}

Eventually, even using shape control, shell dropping typically fails before
the horizons merge: the outflow condition on the
inner boundary of the inner shell fails {\em before} the outflow condition
on the inner boundary of the next shell becomes satisfied.  
This problem occurs because the map ${\cal M}_{\rm E}$ shrinks
the entire grid---including regions inside each hole---faster and faster as
the holes approach each other.
As soon as the velocity of excision boundary towards the center of the
hole becomes large enough, the boundary becomes timelike,
and the excision algorithm fails.  To remedy this, we
turn off shell dropping and we turn on size control. Size control
 slows down the infall of the excision surface (by
pulling the excision sphere towards the horizon), thus keeping
the characteristic speeds from changing sign.  Size control
is implemented by changing the map parameters $\lambda_A^{00}(t)$,
which previously were set to zero,
to values given
by Eqs.~(\ref{eq:lambda-def}) and~(\ref{eq:SizeControlBigLambda00}).
This causes the size of each excision boundary to be driven towards
some fraction $\eta$ of the size of the appropriate horizon.
For the equal-mass runs A, C, and D we use $\eta=0.8$ for
both black holes.  For the unequal-mass run B
and the generic cases E and F
we use $\eta=0.9$
and $\eta=0.7$ for the smaller and larger black holes, respectively.
These values do not require fine-tuning; for instance, case
E still runs through merger and ringdown if we change $\eta$ to
$0.9$ for both black holes.
The main criterion for choosing this parameter is that a grid
sphere of radius $\bar{b}_A$ (see Figure~\ref{fig:DistortionMapGrid}) should
not be mapped outside the cube of side $2\bar{a}_A$.  
If this occurs, the map ${\cal M}_{\rm S}$ becomes singular and the run fails.
Figure~\ref{f:CharSizeControl} 
shows the minimum characteristic speed at particular domain boundaries
for two evolutions of case F: 
one with size control and one without.  Without size control, the outflow
condition on the excision boundary fails at proper separation $1.67 M$
and time $t=127.24M$.
With size control, the evolution proceeds to proper separation $0.015 M$,
time $t=130.17$, well past the formation of a common 
apparent horizon, which occurs at proper separation $1.4M$ (time $t=128.23M$).

\subsubsection{Ringdown}
\label{sec:ringdown}
At some time $t=t_m$ shortly after 
a common apparent horizon forms, we define a new grid, 
composed only of spherical shells.  This grid has only
a single excision boundary inside the common horizon.
We also define a new map ${\cal M}_{\rm ringdown} : \bar{x}^i \to x^i$ that
can be written as a composition of simpler maps:
\begin{equation}
\label{eq:MapCompositionRingdown}
{\cal M}_{\rm ringdown} := {\cal M}_{\rm Tf} \circ
                         {\cal M}_{\rm Ef} \circ
                         {\cal M}_{\rm R}  \circ {\cal M}_{\rm S_3}.
\end{equation}
This is similar to the ringdown maps described in 
Refs.~\cite{Scheel2008,Chu2009} (which had no translation) and 
Ref.\cite{Lovelace:2009} (which had no rotation).
The new translation map ${\cal M}_{\rm Tf}$ is tied to the center
of the new common horizon, and is not continuous with the old translation
map ${\cal M}_{\rm T}$ except near the outer boundary where both translation
maps are the identity.  
The new expansion map ${\cal M}_{\rm Ef}$ is chosen to be
continuous with the old expansion map ${\cal M}_{\rm E}$ at the outer
boundary, but it is the identity near the merged black hole.
This new expansion map ${\cal M}_{\rm Ef}$ smoothly becomes
constant in time shortly after merger.  The rotation map 
${\cal M}_{\rm R}$ is continuous at merger, and after merger it
smoothly becomes constant in time.
The map ${\cal M}_{\rm S_3}$ has the same
form as Eqs.~(\ref{eq:DistortionMap1}--\ref{eq:DistortionMapF})
except the function $\bar{\rho}_{A}(\bar \theta_A,\bar \phi_A)$
defined in Eq.~(\ref{eq:DistortionMaprmax}) is replaced by
$\bar{\rho}_{3}={\rm constant}$, where the constant value corresponds
to a subdomain boundary.

As described in Ref.~\cite{Scheel2008}, at $t=t_m$ the parameters 
$\lambda_A^{\ell m}$ are chosen so that the common apparent horizon
is stationary and
spherical in the grid frame. Similarly, the map ${\cal M}_{\rm Tf}$
is chosen at $t=t_m$ so that the center of the horizon is located at the 
grid-frame origin and is stationary in the grid frame.  

Once we have defined the maps at $t=t_m$,
we then interpolate all variables from the old grid to the new grid.
Note that the grid frame changes discontinuously in time at $t=t_m$.
Because of this, we take care to properly use both the new map 
${\cal M}_{\rm ringdown}$ and the old map ${\cal M}$ in doing this interpolation,
so that the inertial frame and quantities defined in that frame remain
smooth.  In particular, the gauge source function $H_a$ 
is still determined by Eq.~(\ref{eq:QuasiEquilibriumGaugeFalloff}) during
ringdown, and remains smooth at $t=t_m$.  

For $t>t_m$, the parameters
$\lambda_A^{\ell m}(t)$ are determined by a feedback control system that
keeps the common apparent horizon stationary in the grid frame.
Likewise, the map ${\cal M}_{\rm Tf}$ keeps the common apparent horizon
centered at the origin in the grid frame.

\subsection{Results}
\label{sec:results}

%
\begin{figure}
\centerline{\includegraphics[width=3in]{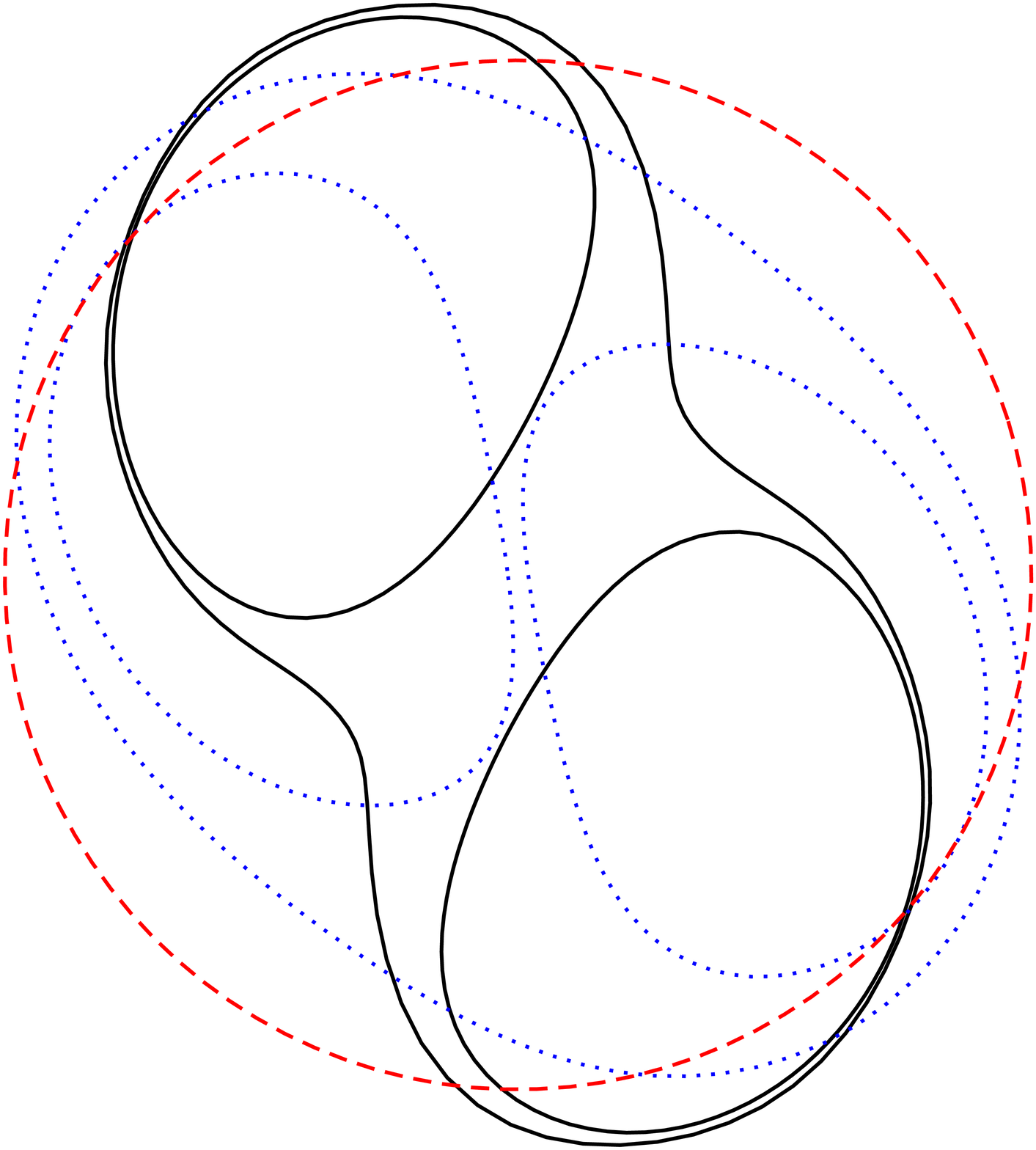}}
\caption{\label{f:AH_equal_mass} Three snapshots of the
  apparant horizons of the non-spinning equal-mass binary black hole
  merger (case A).  Solid curves are the orbital plane cross section
  when a common horizon is first detected; dotted curves represent the
  time of transition between the binary merger and the single hole
  ringdown evolutions; dashed curve shows the final equilibrium
  horizon.}
\end{figure}
\begin{figure}
\centerline{\includegraphics[width=3in]{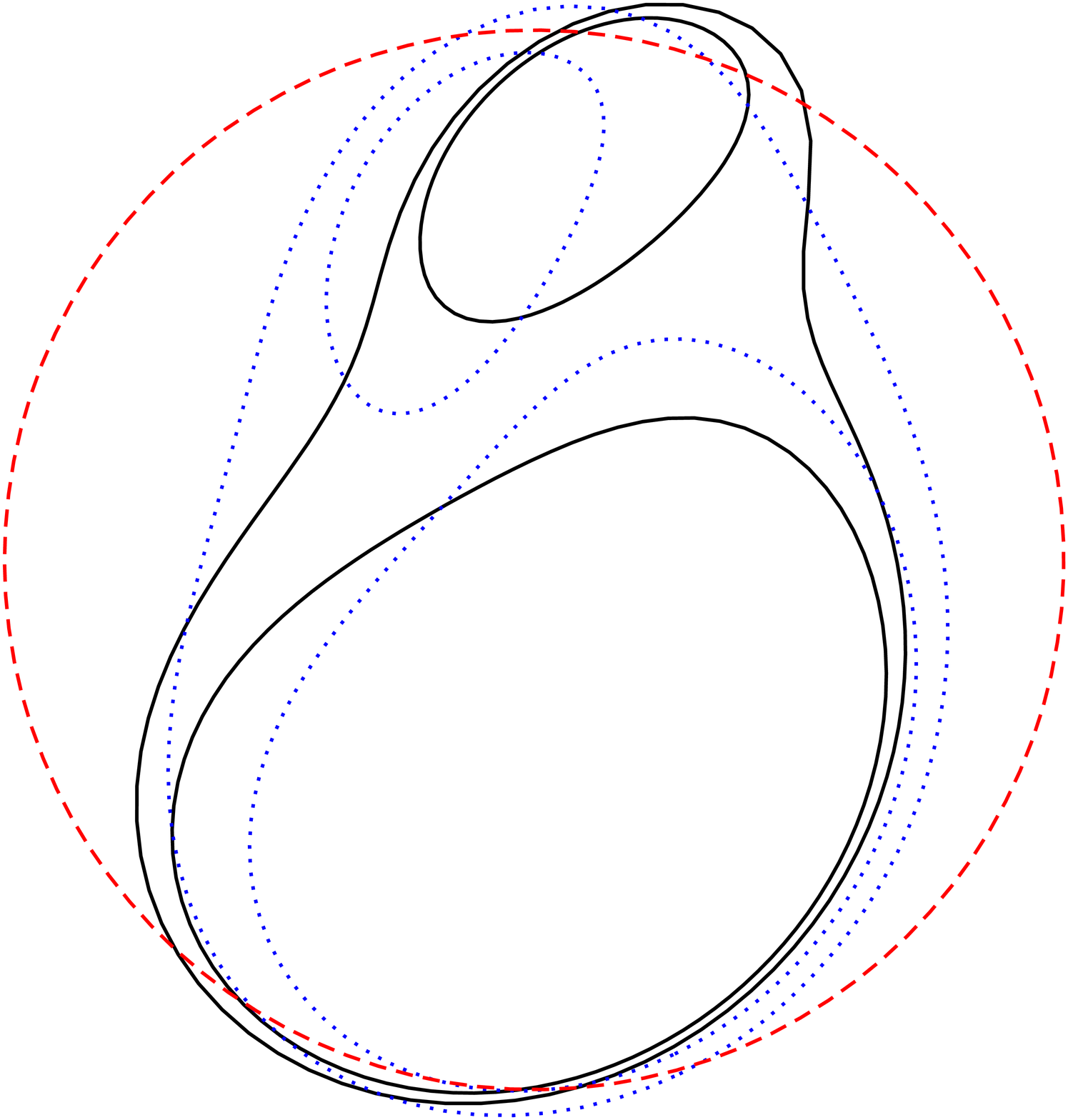}}
\caption{\label{f:AH_2-1_mass} 
  Apparent horizon snapshots as in Figure~\ref{f:AH_equal_mass}, except for
  a 2:1 mass ratio non-spinning binary black
  hole merger (case B).  
}
\end{figure}
\begin{figure}
\centerline{\includegraphics[width=3in]{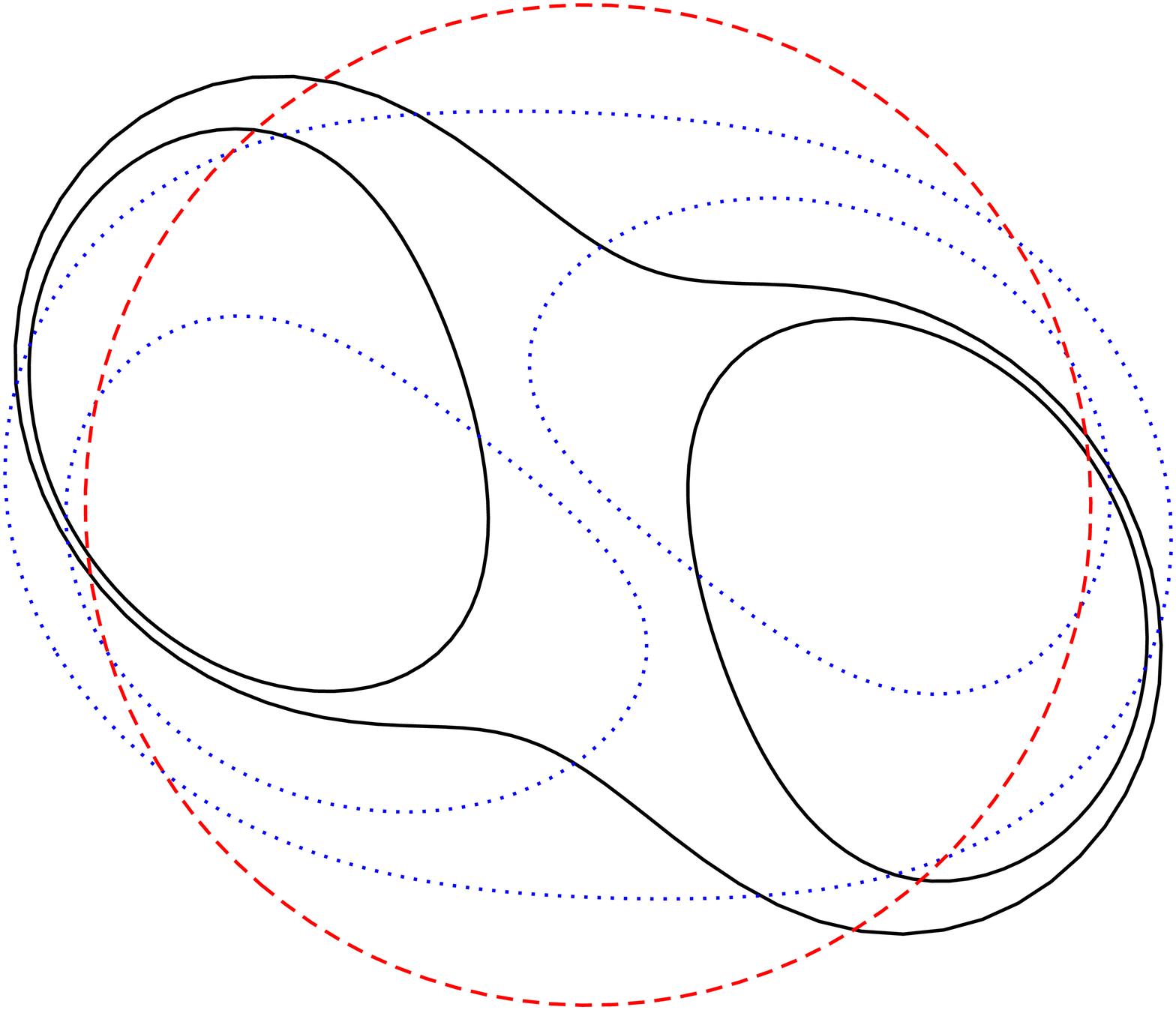}}
\caption{\label{f:AH_anti_aligned} 
  Apparent horizon snapshots as in Figure~\ref{f:AH_equal_mass}, except for
  an anti-aligned-spin equal-mass binary black
  hole merger (case C).  
}
\end{figure}
\begin{figure}
\centerline{\includegraphics[width=3in]{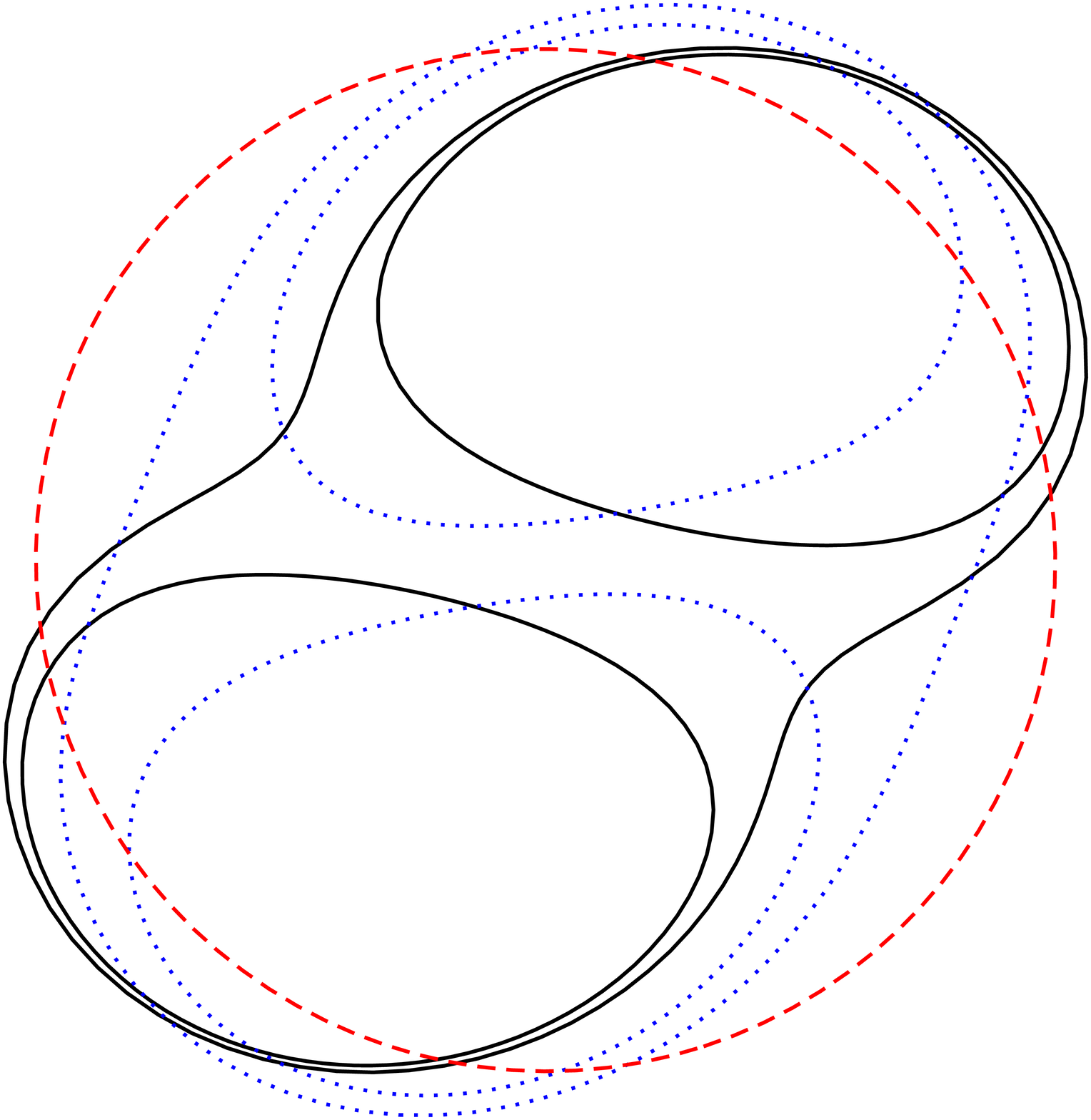}}
\caption{\label{f:AH_aligned_spin} 
  Apparent horizon snapshots as in Figure~\ref{f:AH_equal_mass}, except for
  an aligned-spin equal-mass binary black
  hole merger (case D).  
}
\end{figure}
\begin{figure}
\centerline{\includegraphics[width=3in]{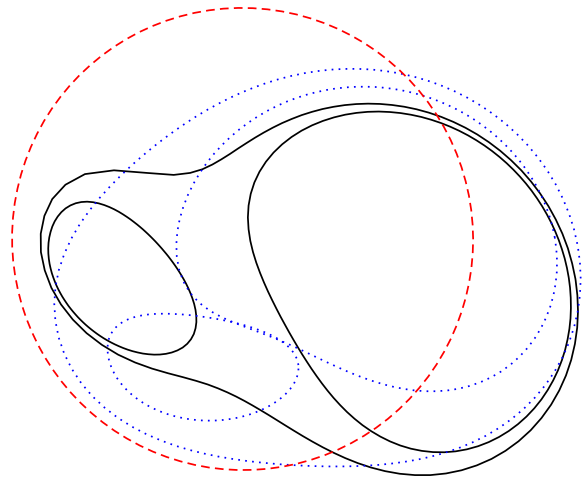}}
\caption{\label{f:AH_generic_long} Three snapshots of the
  apparant horizons of a generic binary black-hole
  merger (case E). The plane of the cross-sections is
  a coordinate plane
  perpendicular to the instantenous orbital axis at the time
  the common horizon is first detected.  The additional degree of
  freedom is fixed by having this plane go through the coordinate-center
  of the shape of the common horizon at its first detection.
  Solid curves are the cross section
  when a common horizon is first detected; dotted curves represent the
  time of transition between the binary merger and the single hole
  ringdown evolutions; dashed curve shows the horizon at a time well
  after merger.}
\end{figure}
\begin{figure}
\centerline{\includegraphics[width=3in]{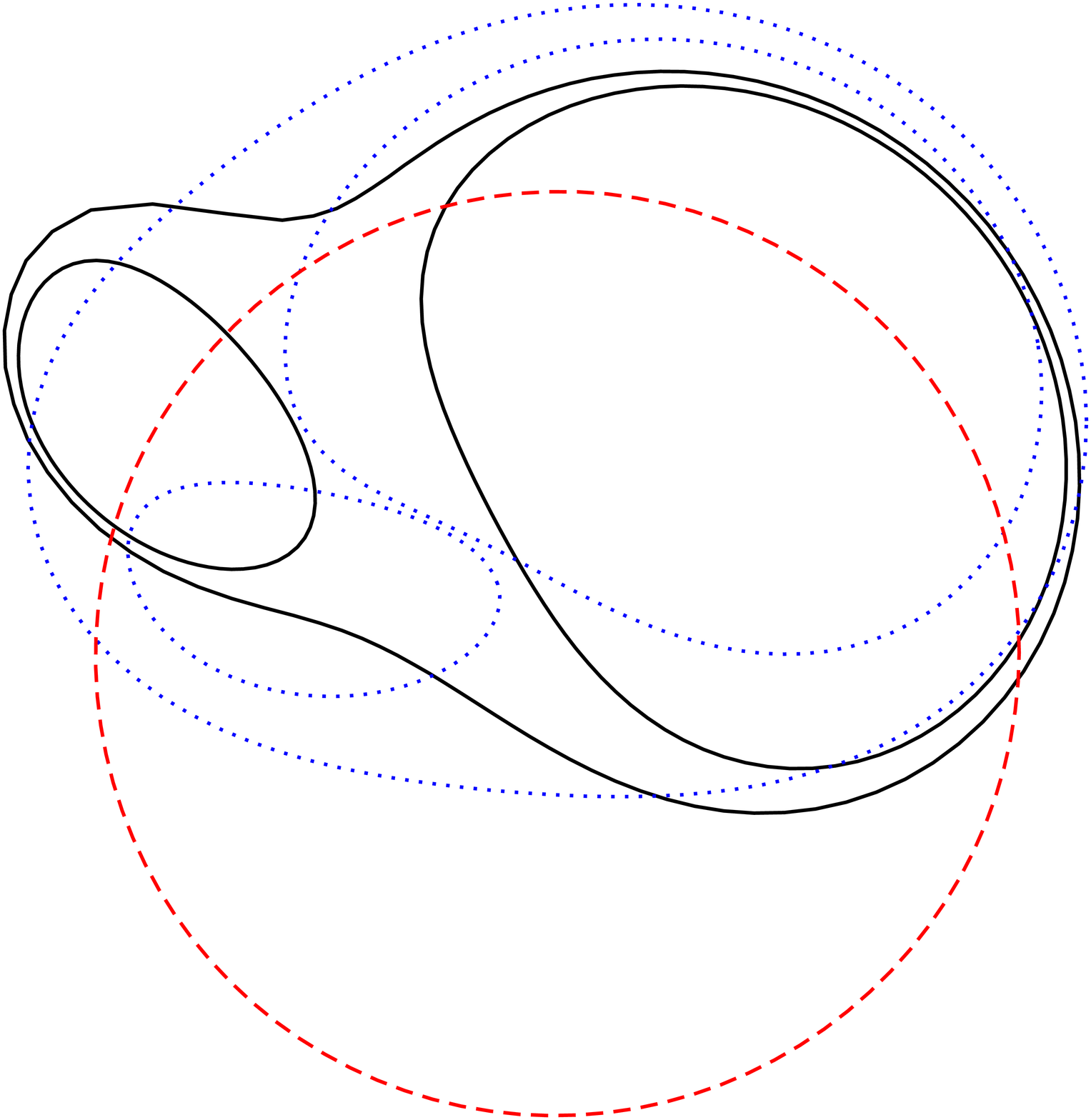}}
\caption{\label{f:AH_generic} 
  Apparent horizon snapshots as in Figure~\ref{f:AH_generic_long}, except for
  the generic binary black-hole merger case F in Table~\ref{tab:Runs}.
}
\end{figure}
\begin{figure}
\centerline{\includegraphics[width=3in]{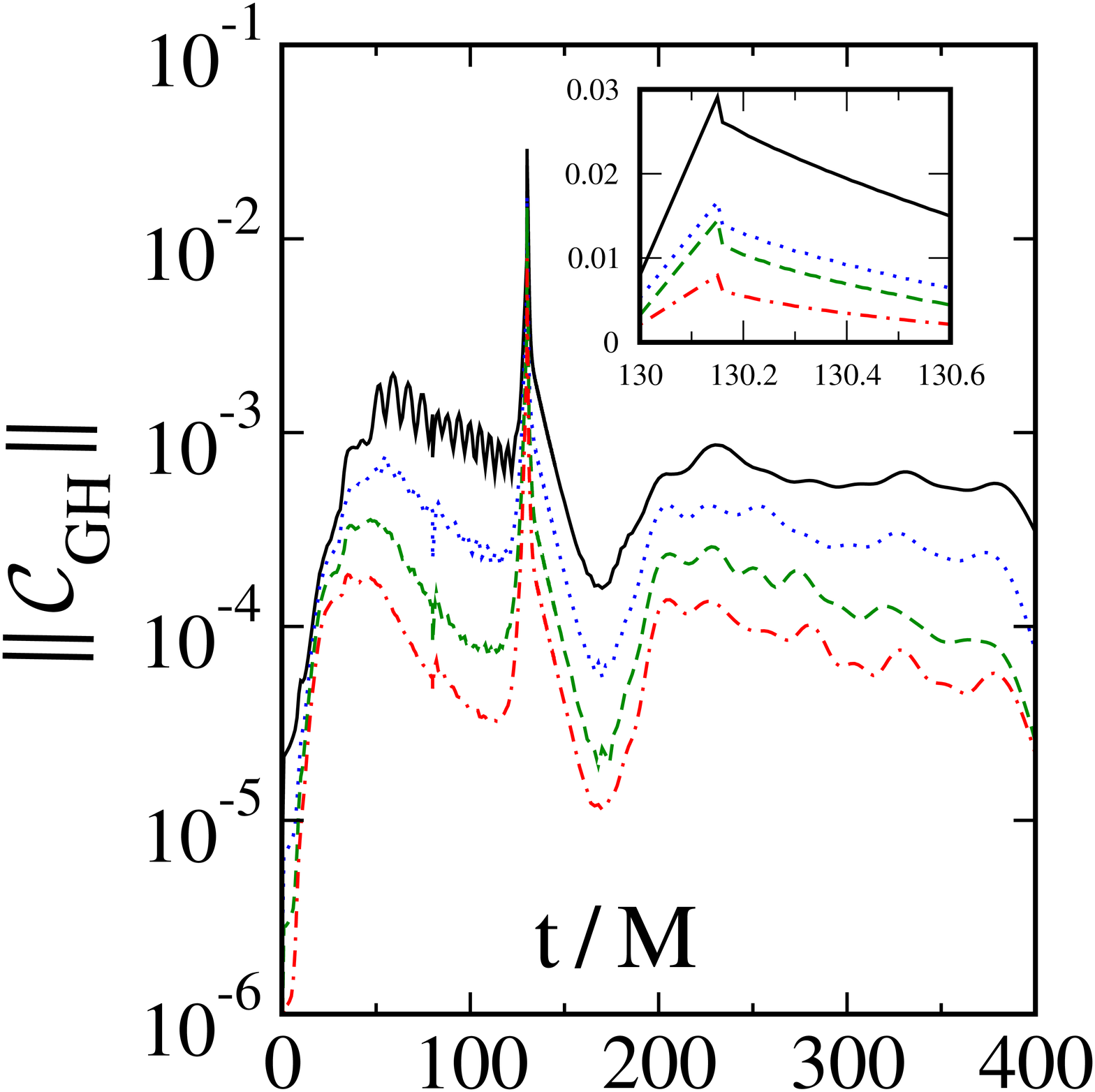}}
\caption{\label{f:GhCeExt} Constraint norm $||{\cal C}_{GH}||$ as a
  function of time for the generic (case F) binary black-hole merger
  and ringdown, computed with four different numerical resolutions.
  The small inset graph shows that numerical convergence is maintained
  (at a reduced rate) even during the most dynamical part of the
  merger. }
\end{figure}

In this section we present some results from the simulations listed in
Table~\ref{tab:Runs}.  We first show snapshots of the
inertial-coordinate shapes of the apparent horizons during merger and
after ringdown.  Figures~\ref{f:AH_equal_mass}, \ref{f:AH_2-1_mass},
\ref{f:AH_anti_aligned}, and~\ref{f:AH_aligned_spin} show
cross-sections of the horizons in the orbital plane for cases A--D; in
these cases, the orbital plane is well-defined and is constant in
time.  For all cases, we show the apparent horizons of the individual
black holes and the common apparent horizon at the time when the
common horizon is first detected (solid curves), and at the time $t_m$ when we
transition to a new grid that has a single excision boundary (dotted
curves).  We also
show the apparent horizon of the final remnant black hole after it
has reached equilibrium (dashed curves).  
Figures~\ref{f:AH_generic_long} and~\ref{f:AH_generic} show cross-sections
of the horizons for cases E and F, in which the orbital plane precesses.
For these cases we show cross-sections in the 
coordinate plane (defined using the flat metric) 
that is perpendicular to the
instantaneous orbital angular 
velocity at the time the common horizon is
first detected, and that passes through the coordinate
center of the common horizon at this time.  We show cross-sections
at three different times: the time of horizon formation, the time $t_m$,
and a time $~300M$ after merger, all with respect to the same plane.
The remnant black hole in cases E and F have nonzero linear momentum both
because of radiation reaction and because of some nonzero linear momentum
present in the initial data; this is why the centers of the
horizons shown in 
Figures~\ref{f:AH_generic_long} and~\ref{f:AH_generic} 
change with time.

For case F, we also show the constraint
norm as a function of time in Figure~\ref{f:GhCeExt}.  We plot
the same quantity $||{\cal C}_{GH}||$ as shown in Figure~\ref{f:NoCubeSpheres}.
This quantity is shown for four numerical resolutions,
and is convergent at all times (although the convergence rate
is smaller near merger when the solution is most dynamical). 
The constraints are largest at $t=t_m$, when we transition
to a grid with a single excision boundary. Just after $t=t_m$ the
constraints decrease discontinuously by a small amount because part of
the computational domain has been newly excised. 
The approximate number of grid points for the four resolutions is
$\left\{ 79^3, 87^3, 95^3, 103^3 \right\}$ just before merger and 
$\left\{ 44^3, 51^3, 57^3, 63^3 \right\}$ during ringdown.


\acknowledgments We thank Jan Hesthaven for helpful discussions.  
We are grateful to Luisa Buchman, Larry Kidder, and Harald Pfeiffer
for the improved kinematical coordinate maps and associated control systems
that handle inspirals of precessing and unequal-mass binaries, 
and we thank Harald Pfeiffer for
help preparing the initial data.  We also thank Luisa Buchman, Tony Chu,
Geoffrey Lovelace,
and Harald Pfeiffer for producing some of the BBH inspiral simulations
that we here extend through merger and ringdown.
We acknowledge use of the Spectral Einstein Code (SpEC).
This work was supported in part by grants from the
Sherman Fairchild Foundation, by NSF grants PHY-0601459, PHY-0652995,
and by NASA grant NNX09AF97G.

\appendix*

\section{Controlling the Maps}
\label{s:control-systems}

The purpose of the shape-control maps ${\cal M}_{\rm S_A}$, with
$\scriptstyle A=\{1,2\}$, is to distort the grid structures around the
black holes so the excision boundaries of the computational domain are
mapped to
surfaces lying just inside and having the same
basic shapes as the apparent horizons.  Choosing the right maps is
equivalent to choosing the right target values for the parameters
$\lambda^{\ell m}_A(t)$ that define these maps, cf.
Eq.~(\ref{eq:DistortionMap3}).  This Appendix describes in some detail
how these parameters are chosen, and what target surfaces are used to
fix these maps in the merger simulations described in
Sec.~\ref{s:MergerRingdownSimulations}.

The grid structures used in our merger simulations,
cf. Sec.~\ref{s:GridStructureFiltering}, have excision boundaries that
are grid-frame coordinate spheres.  The target surfaces ${\cal S}^{\rm
  T}_A$ to which we want to map these grid-frame spheres can be
written in the form
\begin{eqnarray}
\tilde r_A = \sum_{\ell=0}^{\ell_{\rm max}}
                \sum_{m=-\ell}^{\ell} \Lambda^{\ell m}_A(t)
                Y_{\ell m}(\tilde \theta_A,\tilde\phi_A).
\end{eqnarray}
where the expansion coefficients $\Lambda^{\ell m}_A(t)$ define
the target surface and $(\tilde r_A,\tilde\theta_A,\tilde\phi_A)$ are
rest-frame coordinates.

Let $\bar r^{\rm\, target}_A$ denote the radii of the grid-frame
coordinate spheres that are to be mapped onto the target
surfaces ${\cal S}^{\rm T}_A$.  The rest-frame
representations of these coordinate spheres, using
Eqs.~(\ref{eq:DistortionMap1}--\ref{eq:DistortionMap3}), are:
\begin{eqnarray}
\tilde{r}_A      &=& \bar r^{\rm\, target}_A 
                - \sum_{\ell=0}^{\ell_{\rm max}}
                \sum_{m=-\ell}^{\ell} \lambda^{\ell m}_A(t)
                Y_{\ell m}(\tilde \theta_A,\tilde\phi_A).
\end{eqnarray}
Note that the functions $f_A$ that appear in
Eq.~(\ref{eq:DistortionMapF}) are set equal to unity here
because we always choose the target grid-spheres $\bar r^{\rm\,
  target}_A$ to be smaller than the outer radii of the largest
spherical subdomain layers: $\bar b_a \ge \bar r^{\rm\,target}_A$. It
follows that the target spheres will be mapped to the shapes of ${\cal
  S}^{\rm T}_A$ when
\begin{eqnarray}
\lambda^{\ell m}_A(t)= \left\{
\begin{array}{cl}
-\Lambda^{00}_A(t) + \displaystyle
\frac{\bar r^{\rm\,target}_A}{Y_{00}},&\qquad \ell=0, \\
-\Lambda^{\ell m}_A(t),
& \qquad\ell > 0.
\label{eq:lambda-def}
\end{array}
\right.
\end{eqnarray}
These conditions can not be imposed directly for a number of reasons,
but they can be enforced approximately using a feedback control
system.  The control system used in the merger calculations
described in Sec.~\ref{s:MergerRingdownSimulations} is the same one
used to perform our earlier binary black-hole
simulations~\cite{Scheel2006}, so we will not describe it in detail
here.  

To complete the specification of the shape-control maps we must
choose the target grid-spheres 
$\bar r^{\rm\, target}_A$, and the
target surface parameters $\Lambda^{\ell m}_A$.  From Eq.~(\ref{eq:lambda-def})
it follows that the choice of $\bar r^{\rm\, target}_A$ only
effects the $\ell=0$ size control part of the distortion map.
So until size control is activated late in our runs, there is
no need to specify what $\bar r^{\rm\, target}_A$ actually is.
We think of it as being the radius of the suitably scaled apparent
horizon, but this fact does not influence the shape control
part of the map in any way.  Once size control is activated late
in the run, then we require $\bar r^{\rm\, target}_A$
to be the radius of the excision boundary:
\begin{eqnarray}
\bar r^{\rm\, target}_A= 
\bar r^{\rm\, ex}_A.
\label{eq:TargetRdef}
\end{eqnarray}

Next consider the choice of target surfaces ${\cal S}^{\rm T}_A$,
starting with the $\ell>0$ contributions that control the
shape but not the overall scaling of the map.  
The idea is to
choose the target surfaces ${\cal S}^{\rm T}_A$ to be similar in shape
to the apparent horizons ${\cal H}_A$.  The ${\cal H}_A$ can be
represented as smooth surfaces in the rest-frame coordinate system:
\begin{equation}
\tilde r_A
= \sum_{\ell=0}^{\ell_{\rm max}} 
\sum_{m=-\ell}^{\ell} \tilde{S}^{\ell m}_A(t)
Y_{\ell m}(\tilde{\theta},\tilde{\phi}).
\end{equation}
To keep the shapes of the target surfaces similar to the shapes of the
apparent horizons, the target surface parameters $\Lambda^{\ell m}_A$
should be made proportional to $S^{\ell m}_A$.  We find it is
appropriate to scale these coefficients by a factor, $G(\tilde R_A)$, 
which depends on the average radius of the apparent horizon, 
$\tilde R_A=S^{00}_A Y_{00}$:
\begin{eqnarray}
\label{eq:ShapeControlBigLambdalm}
\Lambda^{\ell m}_A(t)=G(\tilde R_A) S^{\ell m}_A(t).
\end{eqnarray}
The larger the apparent horizon radius in relation to the desired
excision radius, the smaller this scaling factor must be to maintain an
appropriate shape for the excision boundry.  In practice, we find the
scale factor
\begin{eqnarray}
\label{eq:ShapeControlGdef}
G(\tilde R_A)= \frac{\bar a_A}{\tilde R_A} \tanh^{\,p}
\left[\left(\frac{\tilde R_A}{\bar a_A}\right)^{\!\!1/p\,}\right],
\end{eqnarray}
works quite well for $p=2$.  
This scaling factor is near unity when
$\tilde R_A<\bar a_A$ and decreases like $1/\tilde R_A$ for larger values of 
$\tilde R_A$. The $\tanh x$ function is introduced
here because it is linear for small values of  $|x|$, and approaches
1 for $x\gg 1$.  The $p$-dependence maintains these
asymptotic forms, but allows
the transition between them to occur more quickly
than the $p=1$ case.

We have found no need to introduce the $\ell=0$ size control parts
these maps, until the final plunge phase just before the black holes
merge.  So during most of our merger simulations we simply set
$\lambda^{00}_A(t)=0$.  This allows the apparent horizons to grow
relative to the grid coordinates as the map ${\cal M}_E$ contracts the
rest-frame relative to the inertial coordinates.  This growth in the
apparent horizons in the grid frame allows us to drop a number of
spherical subdomain layers as the evolution progresses, 
and this helps remove unwanted constraint
violations from the computational domain.  

At some point however, we may not be able to drop additional spherical
subdomain layers: either because we run out of layers,
or because the
inner boundary experiences incoming characteristic fields when the
apparent horizon expands too quickly during the final plunge.  When
either of these conditions occurs we turn on the $\ell=0$ part of the
shape-control maps, to keep the radius of the excision boundary $\bar
r^{\rm\,ex}_A$ close to the size of the apparent horizon.  It will be
useful to define
\begin{eqnarray}
\label{eq:SizeControlDeltardef}
\Delta r_A = \eta \tilde R_A- \bar r^{\rm\,ex}_A.
\end{eqnarray}
We choose $\eta$ so that when $\Delta r_A<0$ there is no need to
impose size control, and use $\eta$ in the range $0.7\le \eta \le 0.9$
for the merger simulations reported in
Sec.~\ref{s:MergerRingdownSimulations}.  When $\Delta r_A$ becomes
positive we want to turn on size control, but we want to do it in a
fairly smooth and continuous way.  This is done
by defining the function:
\begin{equation}
P(\Delta r_A)=
\left\{
\begin{array}{cl}
0,&\Delta r_A<0\\
\tanh\left(\displaystyle
\frac{40 \Delta r_A}{\tilde R_A}\right),&\Delta r_A\ge 0\\
\end{array}
\right.,
\label{eq:SizeControlPdef}
\end{equation}
which vanishes for $\Delta r_A\le 0$, and asymptotically approaches
$P(\Delta r_A)\rightarrow 1$ for $\Delta r_A\gg \tilde R_A/40$.  We
find that an appropriate value for the overall scale
of the target surfaces ${\cal S}^{\rm T}_A$, i.e. their
$\ell=0$ components,
to be
\begin{eqnarray}
\label{eq:SizeControlBigLambda00}
\Lambda^{00}_A = \frac{\bar r^{\rm\,ex}_A + \Delta r_A P(\Delta r_A)}
{Y_{00}}.
\end{eqnarray}
Once size control is activated,
the target radius is given by
Eq.~(\ref{eq:TargetRdef}): $\bar r^{\rm\,target}_A = \bar r^{\rm\,ex}_A$.
This choice implies that the average rest-frame radius of the excision
boundary, $\tilde{R}^{\rm\,ex}_A=\bar
r^{\rm\,ex}_A-\lambda^{00}_AY_{00}$, approaches $\eta \tilde{R}_A$
 when $\Delta r_A\gtrsim R_A/40$.  

To summarize, the appropriate shape and size control of the inner
  grid structures in our binary black-hole merger simulations are
  deterined by the target shape-control surfaces ${\cal S}^{\rm T}_A$,
  whose definitions are given in
  Eqs.~(\ref{eq:ShapeControlBigLambdalm})
  and~(\ref{eq:SizeControlBigLambda00}).  The shape-control map needed
  to distort the grid coordinates to the shape of ${\cal S}^{\rm
    T}_A$, is given in Eq.~(\ref{eq:lambda-def}).  This mapping is
  enforced with an appropriate feedback control system.


\end{document}